\documentclass[10pt,pdflatex,5p]{elsarticle}

\usepackage{amsmath,amssymb,amsfonts}
\usepackage{algorithmic}
\usepackage{graphicx}
\usepackage{subfig}
\usepackage{textcomp}
\usepackage{xcolor}
\usepackage{balance}
\usepackage{url}
\usepackage{comment}
\usepackage{dblfloatfix}
\usepackage{multirow}
\usepackage{xspace}

\journal{Journal of Systems and Software}

\newcommand{\ID}[1]{}
\newcommand{\R}[1]{\textcolor{black}{#1}}
\newcommand{\M}[1]{\textcolor{black}{#1}}
\newcommand{\Ref}[1]{\textsc{#1}}
\newcommand{\Tool}[1]{\textsf{#1}}
\newcommand{\Git}{\Tool{Git}\xspace}
\newcommand{\GitHub}{\Tool{GitHub}\xspace}
\newcommand{\FinerGit}{\Tool{FinerGit}\xspace}
\newcommand{\Historage}{\Tool{Historage}\xspace}
\newcommand{\AURA}{\Tool{AURA}\xspace}
\newcommand{\RMiner}{\Tool{RMiner}\xspace}
\newcommand{\cregit}{\Tool{cregit}\xspace}

\begin{document}

\begin{frontmatter}



\title{On Tracking Java Methods with Git Mechanisms}


\author[OU]{Yoshiki Higo\corref{cor1}}
\ead{higo@ist.osaka-u.ac.jp}

\author[TIT]{Shinpei Hayashi}
\ead{hayashi@c.titech.ac.jp}

\author[OU]{Shinji Kusumoto}
\ead{kusumoto@ist.osaka-u.ac.jp}

\address[OU]{Graduate School of Information Science and Technology, Osaka University, \\Yamadaoka 1--5, Suita, Osaka 565--0871, Japan}
\address[TIT]{School of Computing, Tokyo Institute of Technology, \\Ookayama 2--12--1--W8--71, Ookayama, Meguro-ku, Tokyo 152--8550, Japan}

%\address{}
%\address{}

\cortext[cor1]{Corresponding author}

\begin{abstract}
Method-level historical information is useful in \R{\ID{\#2.8}various} research on mining software repositories such as fault-prone module detection or evolutionary coupling identification.
\M{An existing technique named \Historage converts a \Git repository of a Java project to a finer-grained one.}
In a finer-grained repository, each Java method exists as a single file.
Treating Java methods as files has an advantage, which is that Java methods can be tracked with \Git mechanisms.
\M{The biggest benefit of tracking methods with \Git mechanisms is that it can easily connect with any other tools and techniques build on \Git infrastructure.}
However, \Historage's tracking has an issue of accuracy, especially on small methods.
More concretely, in the case that a small method is renamed or moved to another class, \Historage has a limited capability to track the method.
In this paper, we propose a new technique\R{, \ID{\#2.2}\ID{\#2.7}\FinerGit,} to improve the trackability of Java methods with \Git mechanisms.
We implement \R{\ID{\#2.2}\ID{\#2.7}\FinerGit} as a system and apply it to 182 open source software projects, which include 1,768K methods in total.
The experimental results show that our tool has a higher capability of tracking methods in the case that methods are renamed or moved to other classes.
\end{abstract}


%\item Our technique scored 85\% as max F-measure while the existing one scored 70\%.
%\item Our technique worked well for methods of any size.
%\item Our tool took only short time to build finer-grained repositories even for large ones.

  
\end{frontmatter}
  
  
\section{Introduction}

\begin{comment}
Version control systems provide a variety of useful functionalities for software development.
For example, developers can quickly check differences between given two versions of code, retrieve any past version of code, and merge code that has been developed by plural developers.
Developers also use historical information that is stored in repositories for debugging, understanding code and checking developer's intent of given changes~\cite{codeban2015icsme}.
%Moreover, if developers use hosting service of version control systems such as \GitHub\footnote{\url{https://github.com/}} or Bitbucket\footnote{\url{https://bitbucket.org/}}, they can manage issues and reviews connected to the source code.
Moreover, if developers use a hosting service of version control systems such as \GitHub or Bitbucket, they can manage issues and reviews connected to the source code.
Dabbish et al.\ reported that many activities of software development are supported and recorded by hosting services or version control systems~\cite{dabbish2012cscw}.
\end{comment}

\begin{comment}
Mining software repositories (in short, MSR) is a research field to find useful knowledge from historical information that is stored in repositories of version control systems, issues tracking systems, mailing list archives, and other information sources.
Source code is a primary target in the MSR research field.
For example, identifying fault-prone modules and evolutionary couplings are common topics.
\end{comment}

\R{\ID{\#2.1}
One feature of version control systems is the ability to know file-level change information.}
Thus, it is easy to identify which files were changed in given commits or counting changes for files in a given repository.
However, many approaches in mining software repositories (in short, MSR) require information on finer-grained units such as Java methods or C functions.
If we want to count changes for Java methods, we need to parse source files to identify method positions and then we need to match method positions with changed code positions to identify which methods were changed.
To conduct finer-grained analyses, developers have to implement code/scripts.
Besides, incorrect analysis results will be obtained if the implemented code/scripts include bugs.

Hata et al.\ proposed a technique, \Historage, which enables Java methods to be tracked with \Git mechanisms~\cite{hata2011iwpse}.
\Historage takes a \Git repository of a Java project as its input, and it outputs another \Git repository in which each method gets extracted as a file.
Treating Java methods as files realizes that developers/practitioners can obtain method-level historical information only by executing \Git commands such as \texttt{git-log}.

\begin{figure}[t]
  %\vspace{-0.5em}
  \centering
  \subfloat[\Git repository.]{
   \includegraphics[bb=0 0 345 372,height=5.5cm]{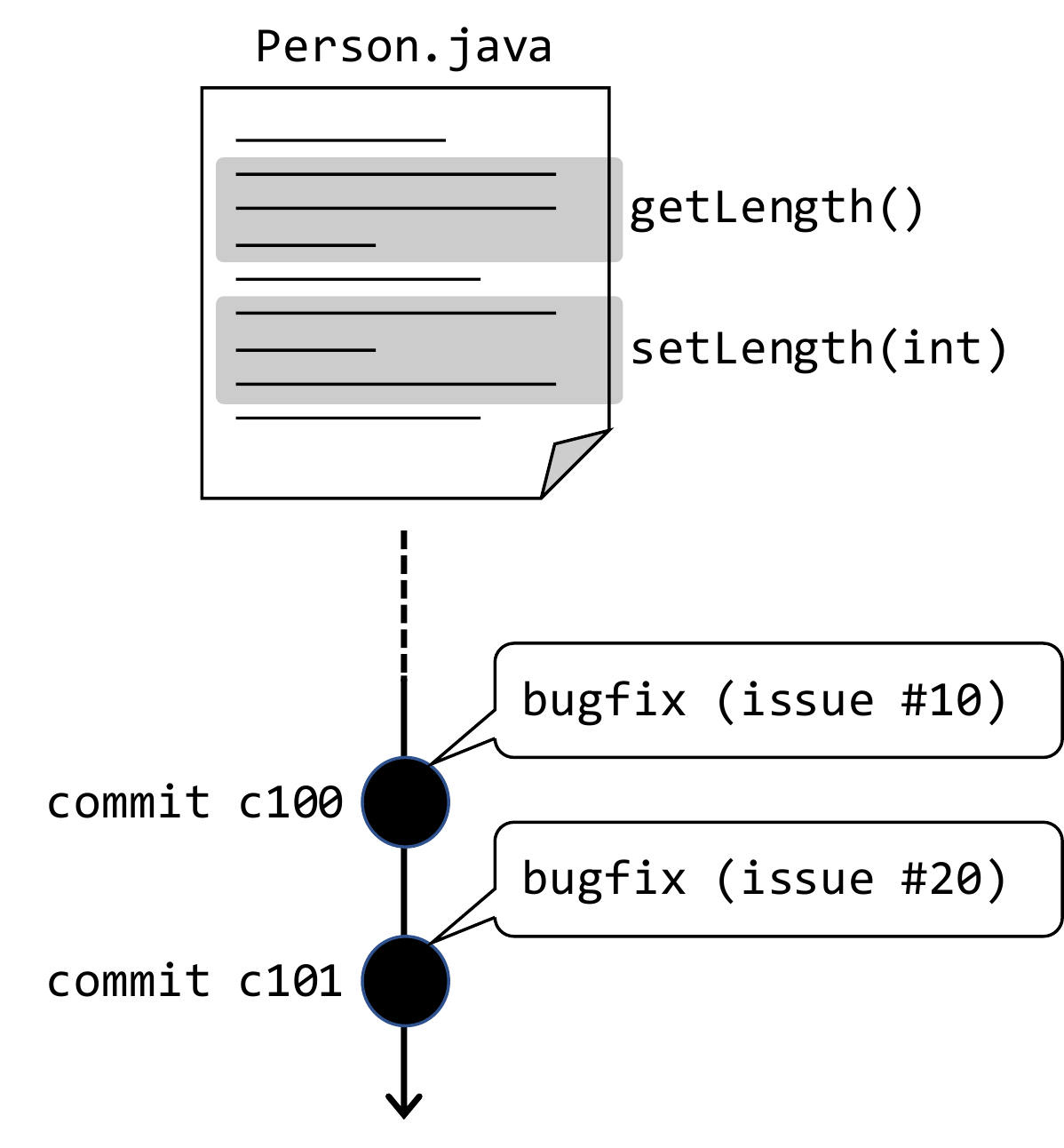}
   \label{fig:GitModel}
  }\\
  \subfloat[\Historage repository.]{
    \includegraphics[bb=0 0 578 372,height=5.5cm]{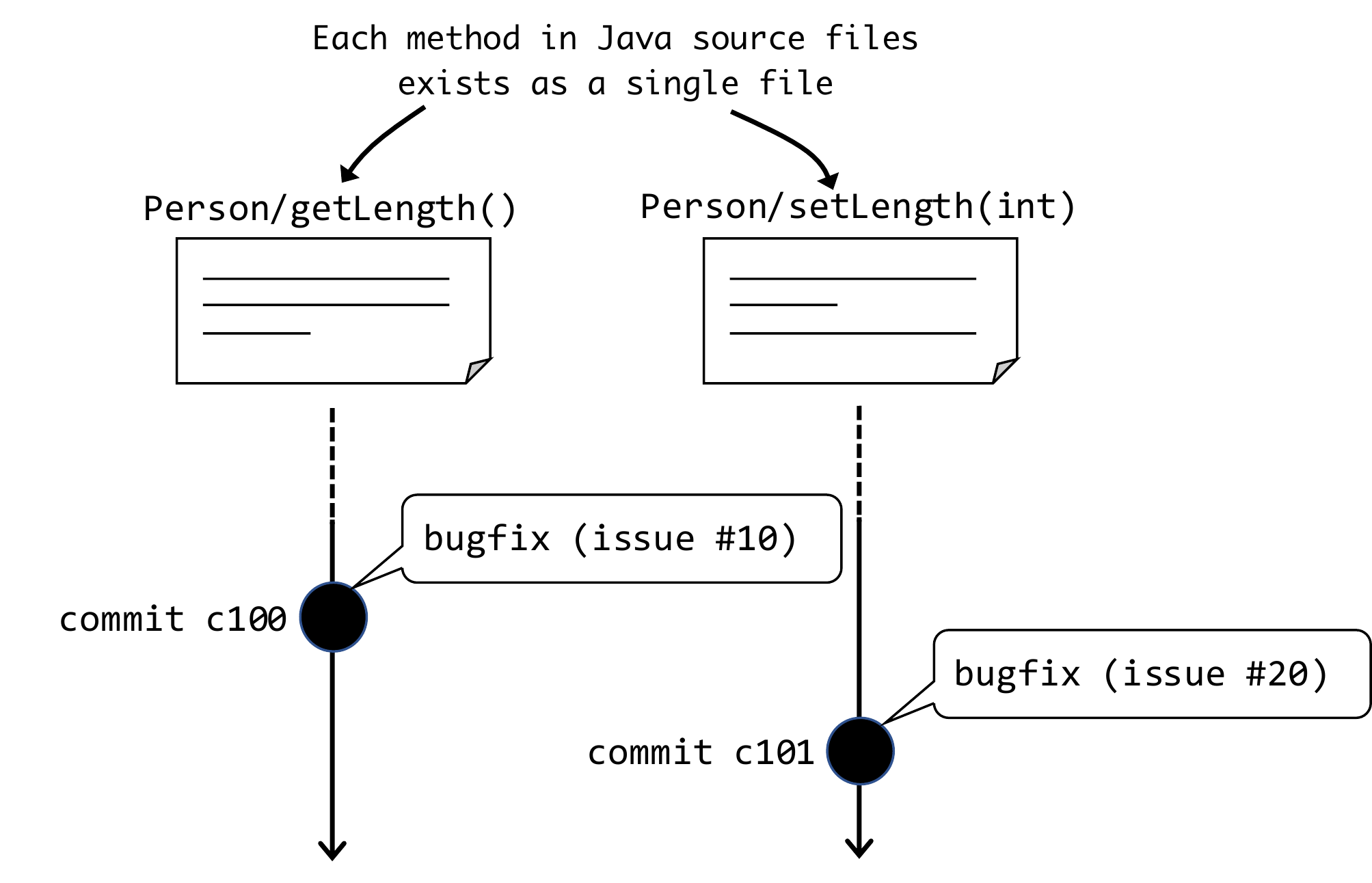}
    \label{fig:HistorageModel}
  }
  \caption{Differences between \Git and \Historage repositories.}
  \label{fig:Model}
  %\vspace{-2.0em}
\end{figure}

Figure~\ref{fig:Model} shows a simple model of \Git and \Historage repositories.
In the \Git repository, file \texttt{Person.java} is managed.
We can see that \texttt{Person.java} was changed in two commits \texttt{c100} and \texttt{c101}.
Information for the changes on \texttt{Person.java} can be retrieved by executing \texttt{git-log}.
However, if we want to know which methods were changed in the two commits, we have to parse \texttt{Person.java} to obtain the positions of the methods and then we have to match method positions with the positions of the changed code in the two commits.
On the other hand, in the \Historage repository, each method exists as a file.
Thus, just executing \texttt{git-log} is sufficient to know in which commits the two methods were changed.
The command identifies that \texttt{getLength()} in \texttt{Person.java} was changed in commit \texttt{c100} and \texttt{setLength(int)} was changed in \texttt{c101}.
%As shown in this example, \Historage enables users to identify method-level information just by executing \Git commands.

\begin{figure}[t]
%  \vspace{-0.5em}
  \centering
  \subfloat[\Git.]{
   \includegraphics[bb=0 0 398 629,width=0.34\textwidth]{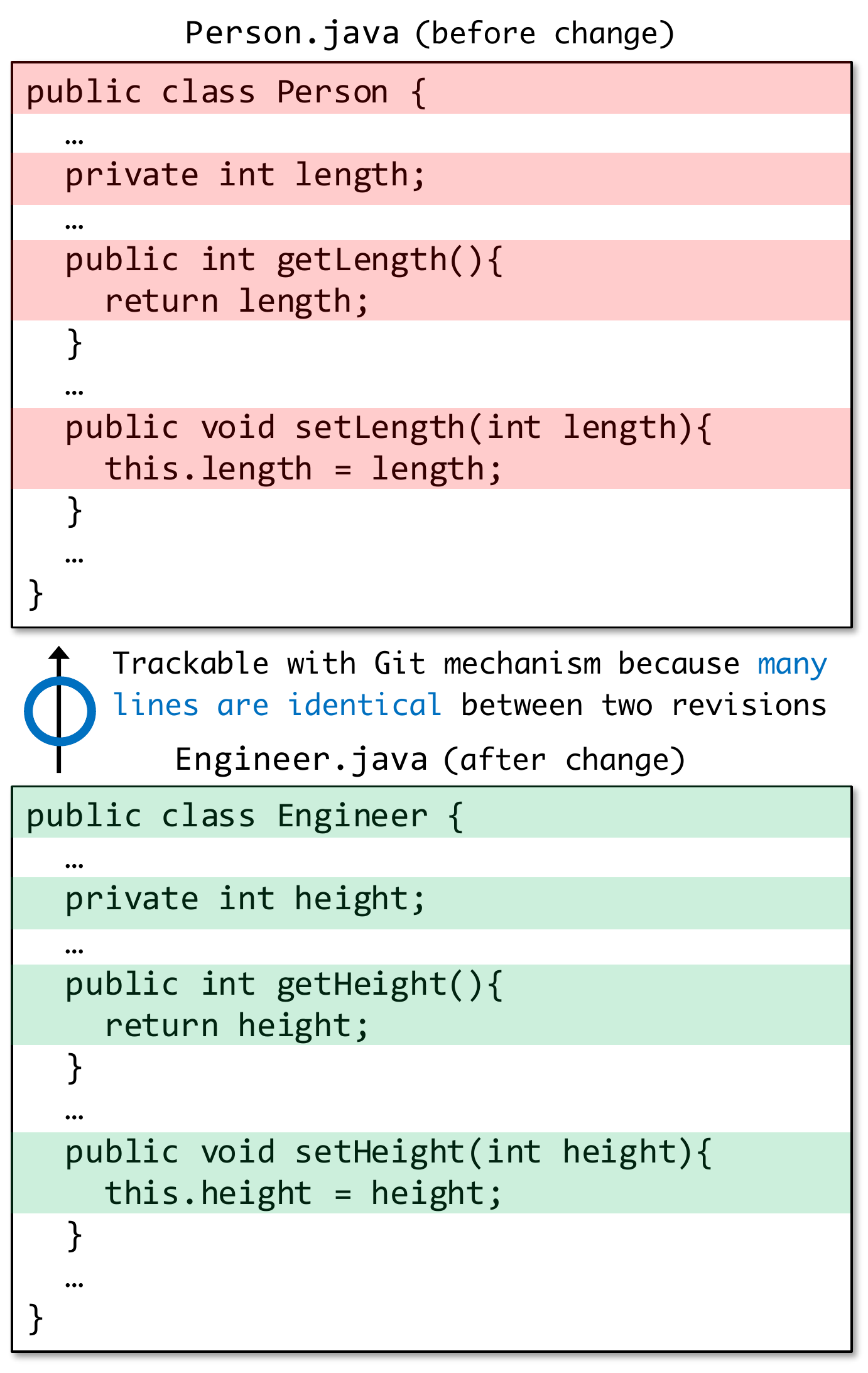}
   \label{fig:GitTrackingModel}
  }\\
  \subfloat[\Historage.]{
    \includegraphics[bb=0 0 395 509,width=0.34\textwidth]{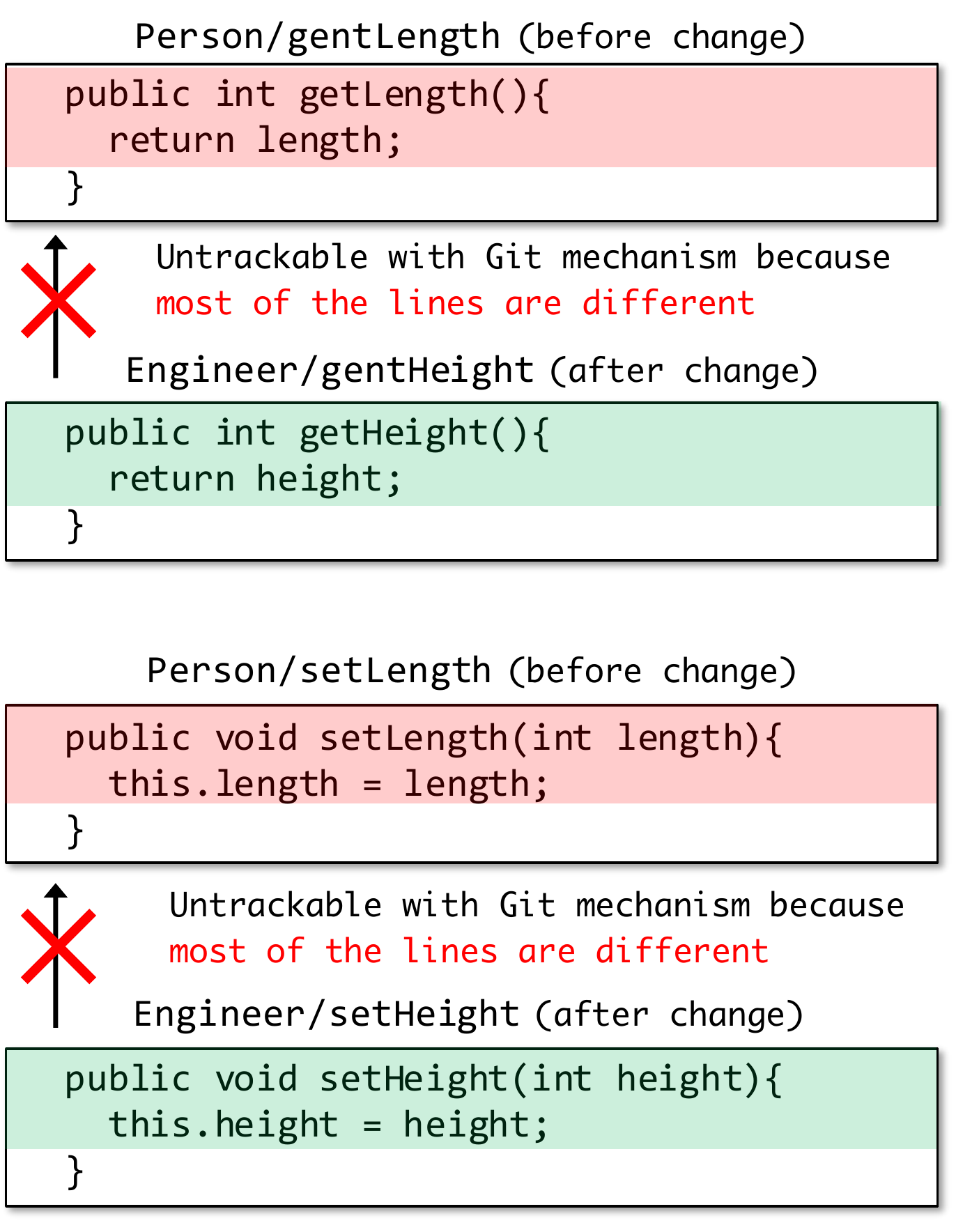}
    \label{fig:HistorageTrackingModel}
  }
  %\vspace{-0.5em}
  \caption{Trackability differences between \Git and \Historage repositories.}
  \label{fig:TrackingModel}
%  \vspace{-1.5em}
\end{figure}

However, \Historage has a limited capability of tracking methods in the case that methods are renamed or moved to other classes.
We explain the issue with Figure~\ref{fig:TrackingModel}, which shows refactorings on file \texttt{Person.java} in Figure~\ref{fig:Model}.
The refactorings include the following four changes.
\begin{description}
  \item[\Ref{Rename Class}:]\texttt{Person} $\rightarrow$ \texttt{Engineer}
  \item[\Ref{Rename Field}:]\texttt{length} $\rightarrow$ \texttt{height}
  \item[\Ref{Rename Method} (Getter):]\texttt{getLength} $\rightarrow$ \texttt{getHeight}
  \item[\Ref{Rename Method} (Setter):]\texttt{setLength} $\rightarrow$ \texttt{setHeight}
\end{description}

In the case of the changes in Figure~\ref{fig:TrackingModel}\subref{fig:GitTrackingModel}, the \Git rename detection function can identify that file \texttt{Person.java} was renamed to \texttt{Engineer.java} because the two files sufficiently share the identical lines.
On the other hand, in the \Historage repository, files of Java methods get much smaller than their original file as shown in Figure~\ref{fig:TrackingModel}\subref{fig:HistorageTrackingModel}.
Thus, the ratio of the changed lines against all the lines gets higher, which makes the \Git function not work well.

Hata et al.\ addressed that changing the threshold for the \Git rename function is a way to realize a better method tracking~\cite{hata2011iwpse}.
They recommend using 30\% instead of 60\%, which is a default value of \Git.
However, we consider that only using a lower threshold may produce incorrect tracking results.
For example, if we use 30\% instead of 60\%, the \Git rename function can identify that \texttt{Engineer/getHeight()} is a renamed file of \texttt{Person/getLength()}.
However, at the same time, \texttt{Person/getLength()} can be tracked wrongly from \texttt{Engineer/setHeight(int)} because their similarity is 1/3, which is higher than 30\%.

Tracking method accurately is essential.
If not, MSR approaches using historical data gets affected.
Hora et al.\ reported that between 10 and 21\% of changes at the method level in 15 large Java systems were untracked in the context of refactoring detection~\cite{hora2018icse}.
They also found that 37\% of the top-25\% most changed entities (classes and methods) have at least \R{\ID{\#2.8}one untracked change} in their histories.
By assessing two MSR approaches, they detected that their results could be improved when untracked changes were resolved.

In this paper, we propose a new technique \R{named \FinerGit} to improve the trackability of Java methods.
Several research areas benefit from \R{\FinerGit}.
\R{\FinerGit} is useful for studies in the context of assessing bug introducing changes~\cite{kim2006ase,kim2008tse,sliwerski2005msr} or detecting code authorship~\cite{meng2013icsm,rahman2011icse}.
More broadly, any study that compares two versions of methods can be benefited, for example, API evolution detection~\cite{kim2011icse,soares2010software}, code warning prioritization~\cite{balachandran2013icse,kim2007esecfse}, and many other.

\begin{figure}[t]
  %\vspace{-1.0em}
  \centering
  \includegraphics[bb=0 0 419 764,width=0.37\textwidth]{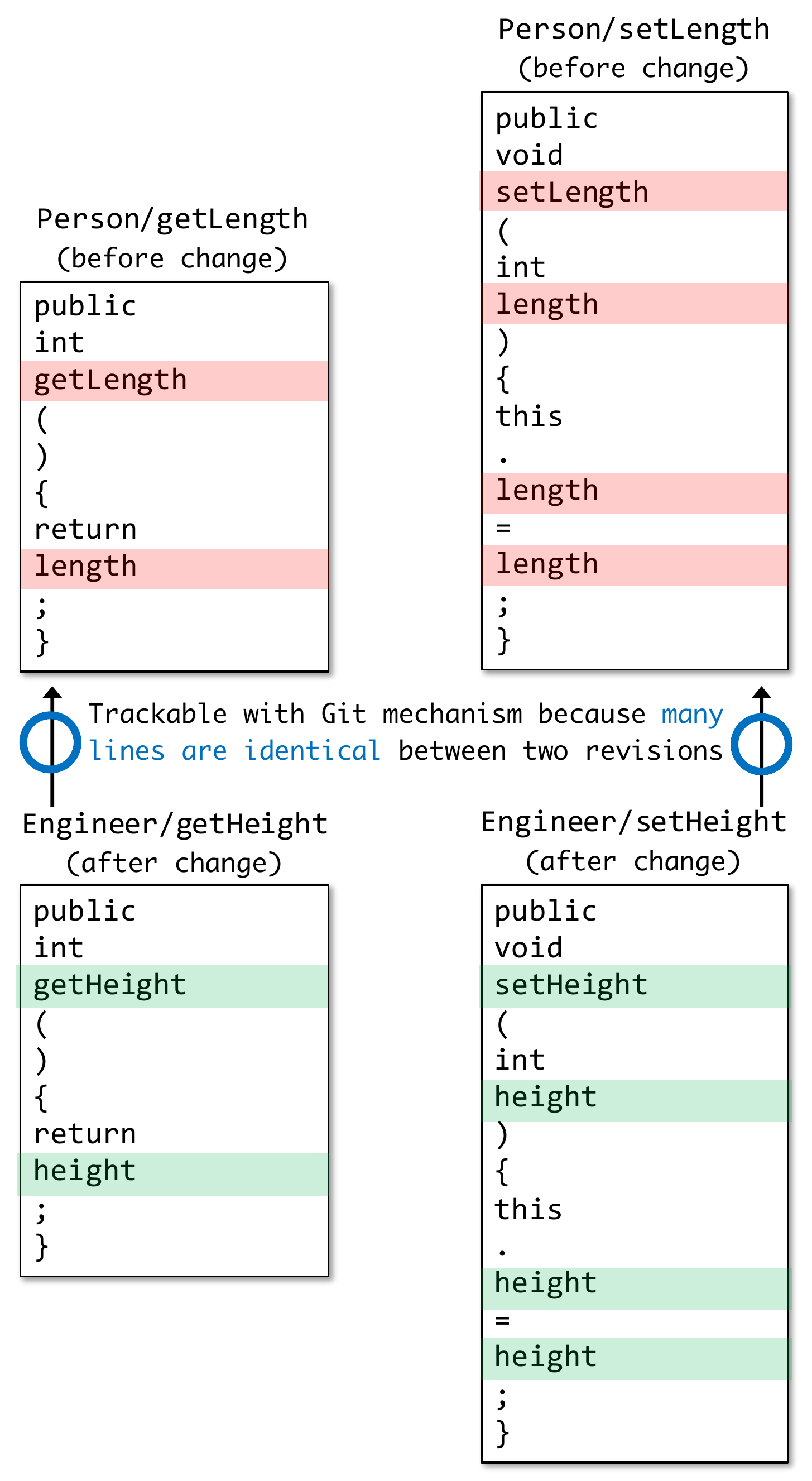}
  \caption{Tracking files with our technique.}
  \label{fig:FinerGitTrackingModel}
  %\vspace{-1.0em}
\end{figure}

The main contributions of this paper are the followings.
\begin{itemize}
  \item We raise an issue on method trackability in \Historage.
  \item We propose a new technique\R{, \FinerGit,} to increase method trackability with \Git mechanisms.
  \item We provide a software tool based on \R{\FinerGit}. The tool is open to the public on \GitHub\footnote{\url{https://github.com/kusumotolab/FinerGit}}. The tool is sufficiently fast even for huge repositories, as shown in the evaluation.
  \item We show the experimental results on the tracking results of 182 open source software~(OSS) projects.
  These experiments have two aspects.
  \R{First, they clarify the advantage of \FinerGit with an existing technique, \Historage.}
  Second, they are the first attempt of large-scale empirical studies for the tracking results of method-level repositories.
\end{itemize}

The remainder of this paper is organized as follows:
\R{in Section~\ref{sec:motivation}, we explain our research goal and our key idea to achieve the goal;
in Section~\ref{sec:proposal}, we propose our new technique named \FinerGit on the top of the key idea;}
Section~\ref{sec:implementation} describes an implementation of \R{\FinerGit}; 
then, we report the evaluation results with the implementation in Section~\ref{sec:evaluation}; 
we also describe threats to validity on the experiments in Section~\ref{sec:validity}; 
related work is introduced in Section~\ref{sec:relatedwork}; 
lastly, we conclude this paper in Section~\ref{sec:conclusion}.

\section{\ID{\#1.1}\ID{\#1.3}\ID{\#2.5}Basic Approach}
\label{sec:motivation}

At present, there are various techniques of tracking source code entities~\cite{dig2006ecoop,godfrey2005tse,kim2005wcre,wu2010icse}. 
Those techniques utilize many types of information such as text similarities, data dependencies, and call dependencies.
On the other hand, in this research, we utilize only line-based text similarity to track Java methods.
The reason is that our research goal is realizing accurate method tracking with \Git mechanisms.

\begin{comment}
In \Git, only line-based text similarities are utilized for identifying renaming and copying of files.
More concretely, \Git performs file comparisons by using hash values.
If the size of a line is equal to or shorter than 64 bytes, a hash value is calculated from the entire line.
If the size of a line is longer than 64 bytes, the line is chunked by 64 bytes, and a hash value is calculated from each chunk.
Thus, even if just a single token in a given line (which is short than 64 bytes) has been changed,
\Git regards that the entire line has been changed.
\end{comment}

The biggest benefit of tracking methods with \Git mechanisms is that it can easily connect with any other tools and techniques built on \Git infrastructure. For example, the following analyses can be easily performed by using the basic commands provided by \Git.
\begin{itemize} 
  \item We can know how many times each method was changed in the past by \texttt{git-log}.
  \item We can know how many developers changed a specified method in the past by collecting author names of the commits in which the method was changed.
\end{itemize}

\Git performs file comparisons by using hash values.
If the size of a line is equal to or shorter than 64 bytes, a hash value is calculated from the entire line.
If the size of a line is longer than 64 bytes, the line is chunked by 64 bytes, and a hash value is calculated from each chunk.
Thus, even if just a single token in a given line (which is shorter than 64 bytes) has been changed,
\Git regards that the entire line has been changed.

Method-level tracking with \Git mechanisms can be realized by treating each method as a single file (a \emph{method file} hereafter).
Based on this idea, Hata et al.\ developed technique named \Historage~\cite{hata2011iwesep}.
However, as explained with Figure~\ref{fig:TrackingModel}, simple extraction as files are inadequate for small methods.
In this research, we propose a file format that each line includes only a single token.
By using this format, each hash is calculated from a single token.
In Figure~\ref{fig:TrackingModel}\subref{fig:HistorageTrackingModel}, \Git regards that the two red lines of methods \texttt{getLength} and \texttt{setLength} were changed, though only the method name and the field name were changed in methods.
As a result, the ratio of unchanged lines becomes 1/3, which is less than 60\% of \Git's default value so that the method is not tracked with \Git mechanisms.

We state two restrictions for the techniques to improve method tracking with \Git mechanisms as follows.
\begin{itemize}
  \item Since the file tracking mechanism in \Git is based on line-based text similarity, the characteristics of methods to be used in comparison must be represented as a sequence of text lines.
        Based on this restriction, complex comparison techniques of file contents such as \textit{tf}/\textit{idf} are not applicable.
  \item Since the contents of method files are visible and are utilized by developers, they should follow a representation of source code in an understandable way by users.
        Users may apply \texttt{git-diff} command to a method file to see how a method was modified, and the obtained difference should represent the difference of method contents in this case.
        Based on this restriction, converting method contents to a sequence of computed numeric values used only for a comparison purpose is not suitable.
\end{itemize}

Figure~\ref{fig:FinerGitTrackingModel} shows how the changes in Figure~\ref{fig:TrackingModel}\subref{fig:HistorageTrackingModel} are treated in \R{\FinerGit}.
The file changing mechanism in this technique satisfies the above restrictions.
The ratio of unchanged lines becomes 8/10 for \texttt{getLength} and 11/15 for \texttt{setLength}.
Both values are higher than 60\%, so that both methods are tracked with \Git mechanisms.

\section{Proposed Technique}
\label{sec:proposal}

Herein, we explain our proposed technique \R{named \FinerGit} to realize a better method tracking with \Git mechanisms.
\M{\FinerGit is designed on the top of the basic approach explained in Section~\ref{sec:motivation}.
\FinerGit consists of (1) naming convention and (2) two heuristics.}

\subsection{Naming Convention}

In \R{\FinerGit}, a file name for a Java method includes the following information:

\begin{itemize}
  \item a class name including the method,
  \item access modifiers of the method,
  \item a return type of the method,
  \item a name of the method, and
  \item a list of parameter types of the method.
\end{itemize}

For example, the file name for method \texttt{setLength} in Figure~\ref{fig:TrackingModel} becomes as follows.

\vspace{0.5em}
\verb|Person#public_void_setLength(int).mjava|
\vspace{0.5em}

\M{Extension \texttt{.mjava} means that this is a method file and the file includes source code of a Java method.}
Including the above information in the file name reflects code changes around a given method as follows.

\begin{itemize}
  \item If the name of the class including the given method is changed,
  the file name of the given method gets changed, but its contents are not changed.
  \item If another method in the class including the given method is changed,
  neither file name nor contents of the given method are changed.
  \item If the signature of the given method is changed,
  the file name of the given method gets changed \M{and its contents are also slightly changed since the contents include the tokens of the method signature.}
  \item If the contents of the given method are changed,
  the file name of the given method does not get changed while its contents get changed.
\end{itemize}

We can track methods with \Git mechanisms in any of the above cases if either of them occurs alone.
However, if a signature of a method is changed and its contents are also changed broadly, it is difficult to track the method.

\begin{figure}[t]
%  \vspace{-1.0em}
  \centering
  \subfloat[\R{\ID{\#2.8}}\Historage.]{
   \includegraphics[bb=0 0 404 321,width=0.35\textwidth]{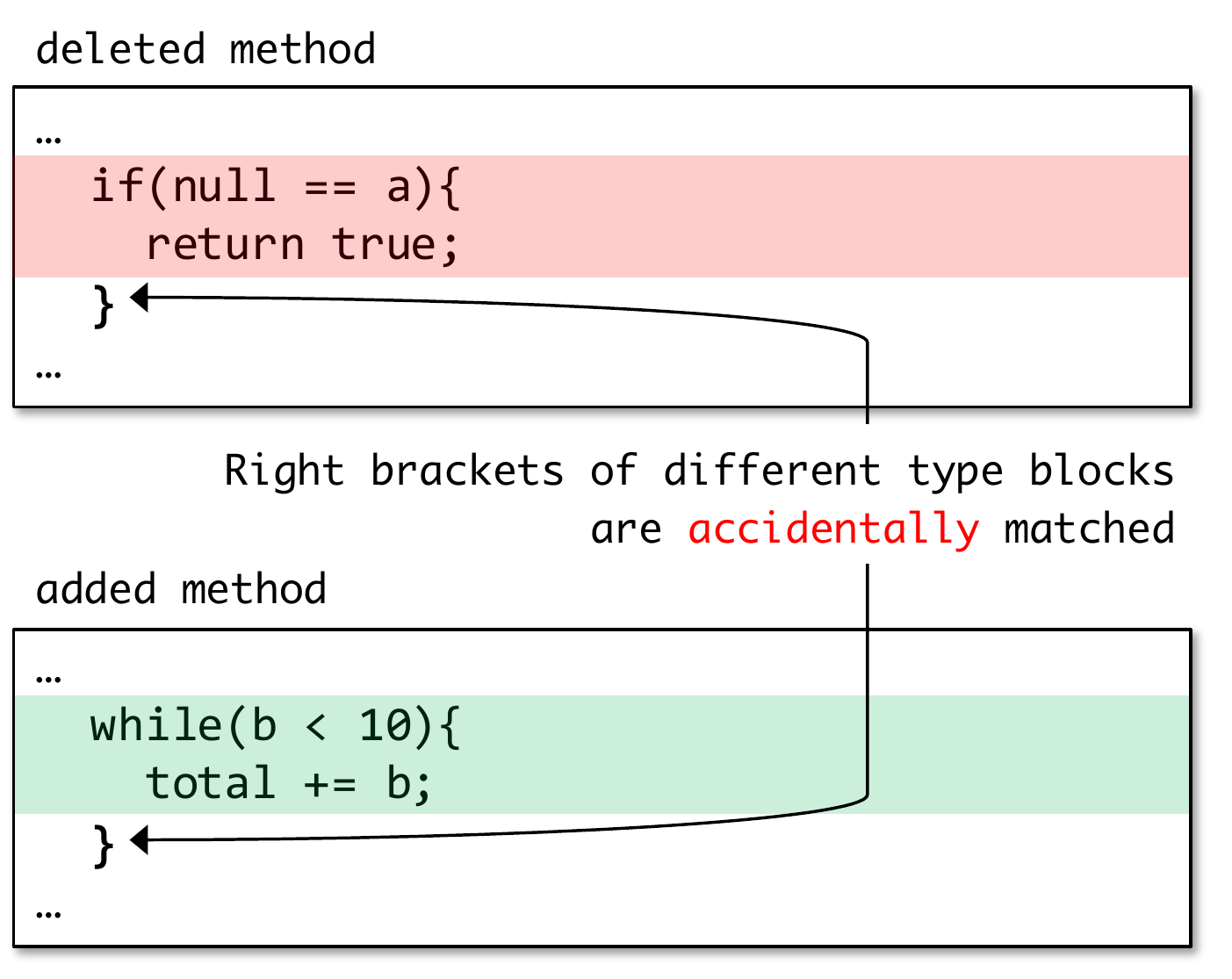}
   \label{fig:Heuristic1Historage}
  }\\
  \subfloat[w/o Heuristic-1.]{
    \includegraphics[bb=0 0 206 658,width=0.1782\textwidth]{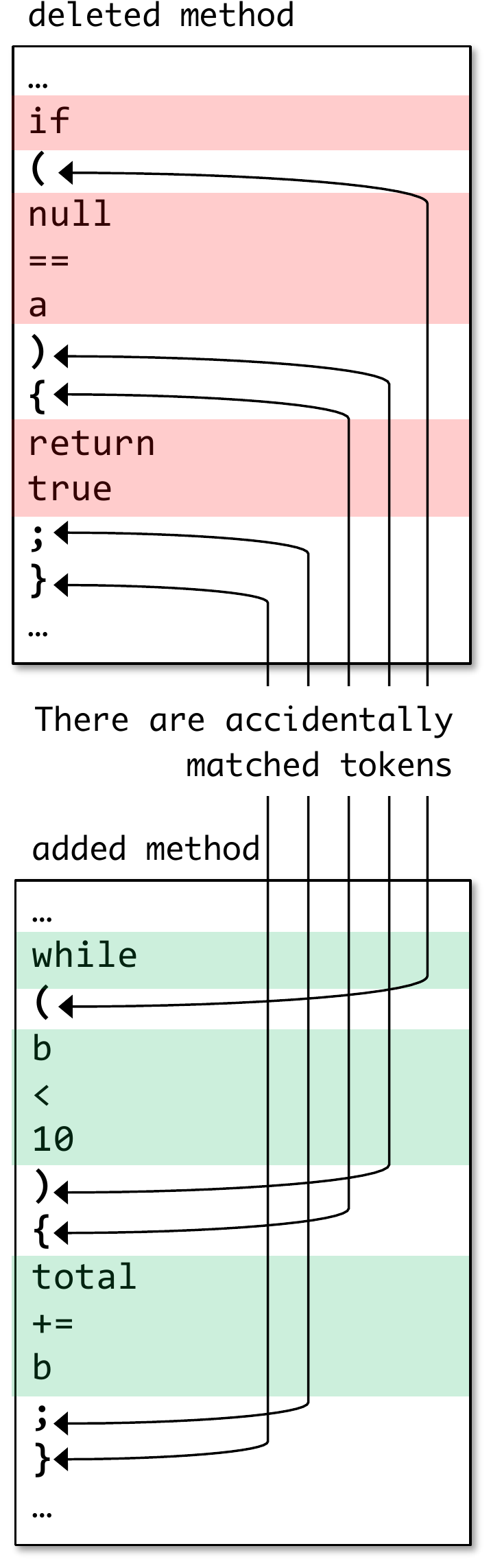}
    \label{fig:Heuristic1Without}
  }
  \subfloat[w/ Heuristic-1.]{
    \includegraphics[bb=0 0 263 657,width=0.228\textwidth]{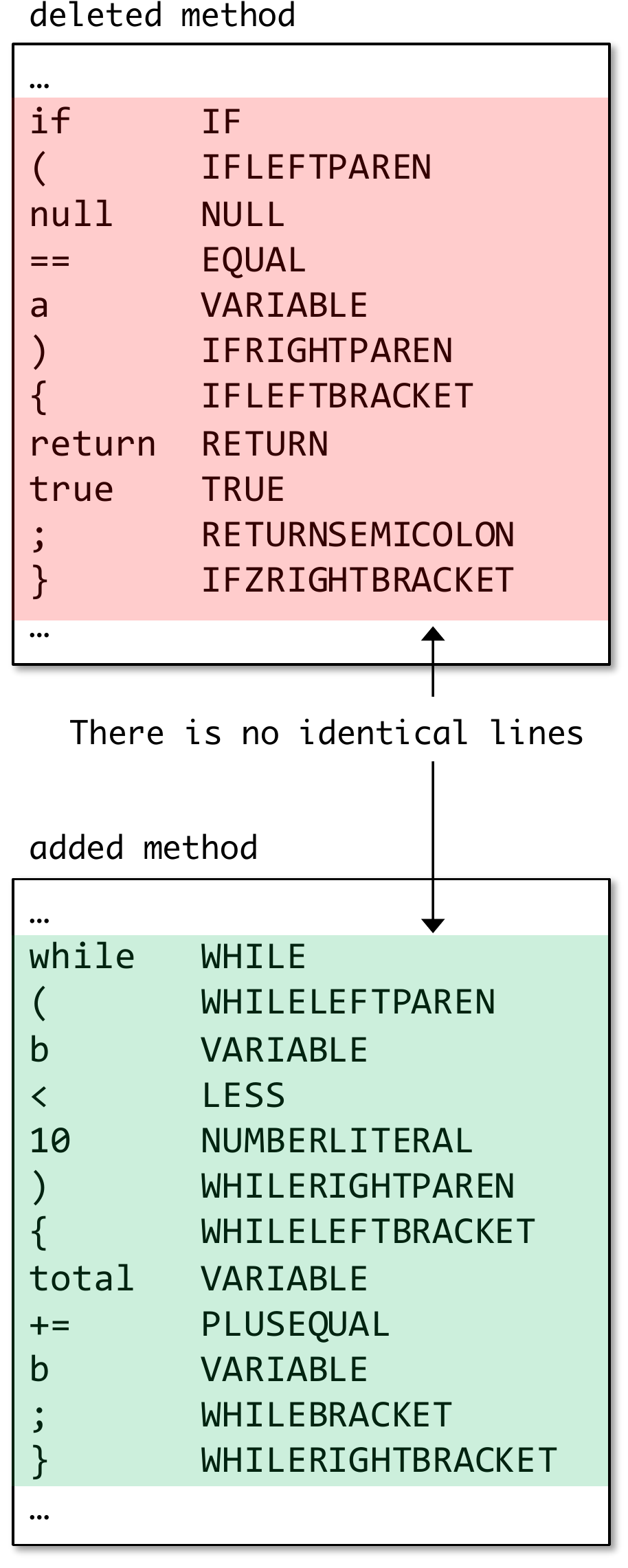}
    \label{fig:Heuristic1With}
  }
  \caption{Tracking files w/o and w/ Heuristic-1.}
  \label{fig:Heuristic1}
%  \vspace{-1.0em}
\end{figure}

\subsection{Introducing Heuristics}
\label{sec:heuristics}

It is not difficult to imagine that \Git tracks wrong methods with \R{\FinerGit} because each line has only a single token and such lines will coincidentally match with many other lines.
Thus, we introduce two heuristics to reduce such coincidental matches of unrelated lines.
\begin{description}
  \item[\textbf{Heuristic-1:}] Classifying brackets, parentheses, and semicolons of termination characters in detail.
  \item[\textbf{Heuristic-2:}] Removing tokens existing in all methods from the targets of similarity calculation.
\end{description}

%\vspace{1.0em}
%\textbf{Heuristic-1}\\
\subsubsection{Heuristic-1}
Some termination characters such as brackets, parentheses, and semicolons are omnipresent in Java source code.
Such termination characters are used as a part of various program elements.
For example, brackets (``\texttt{\{}'' and ``\texttt{\}}'') are used to initialize arrays in addition to code blocks such as if-statements and for-statements.
Thus, if just a bracket is placed on a line, brackets of different roles are coincidentally matched with each other.
Such accidental matchings make the similarity between deleted and added methods inappropriately higher.
To prevent such accidental matchings, we classify termination characters in detail.
More concretely, we add a token explanation to each line.
Token explanations prevent accidental matchings of different-role characters from being matched.
In this heuristic, semicolons, brackets, and parentheses are classified into 18, 21, and 20 categories, respectively.

Figure~\ref{fig:Heuristic1} shows how Heuristic-1 affects method tracking.
Figure~\ref{fig:Heuristic1}\subref{fig:Heuristic1Historage} is a method file that \Historage outputs.
The deleted method includes an if-statement for checking whether variable \texttt{a} is \texttt{null} or not.
The added method includes a while-statement for adding variable \texttt{b} to variable \texttt{total} repeatedly.
Those are different methods, which means a lower similarity between them is better.
In the case of \Historage, the last line of the if-statement coincidentally matches with the last line of the while-statement so that the similarity between them becomes 1/3 (=33\%).
In the case of \R{\FinerGit} without Heuristic-1, the parentheses and the brackets of the if-statement coincidentally matches with ones of the while-statement.
Moreover, the semicolon of the return-statement coincidentally matches with the one of the expression-statement.
As a result, the similarity between them becomes 5/12 (=42\%).
If we introduce Heuristic-1 to this example, the parentheses, the brackets, and the semicolons get unmatched.
Thus, the similarity between them becomes 0/12 (=0\%).

\begin{figure}[t]
  %\vspace{-1.5em}
  \centering
  \subfloat[w/o Heuristic-2.]{
    \includegraphics[bb=0 0 296 623,width=0.23\textwidth]{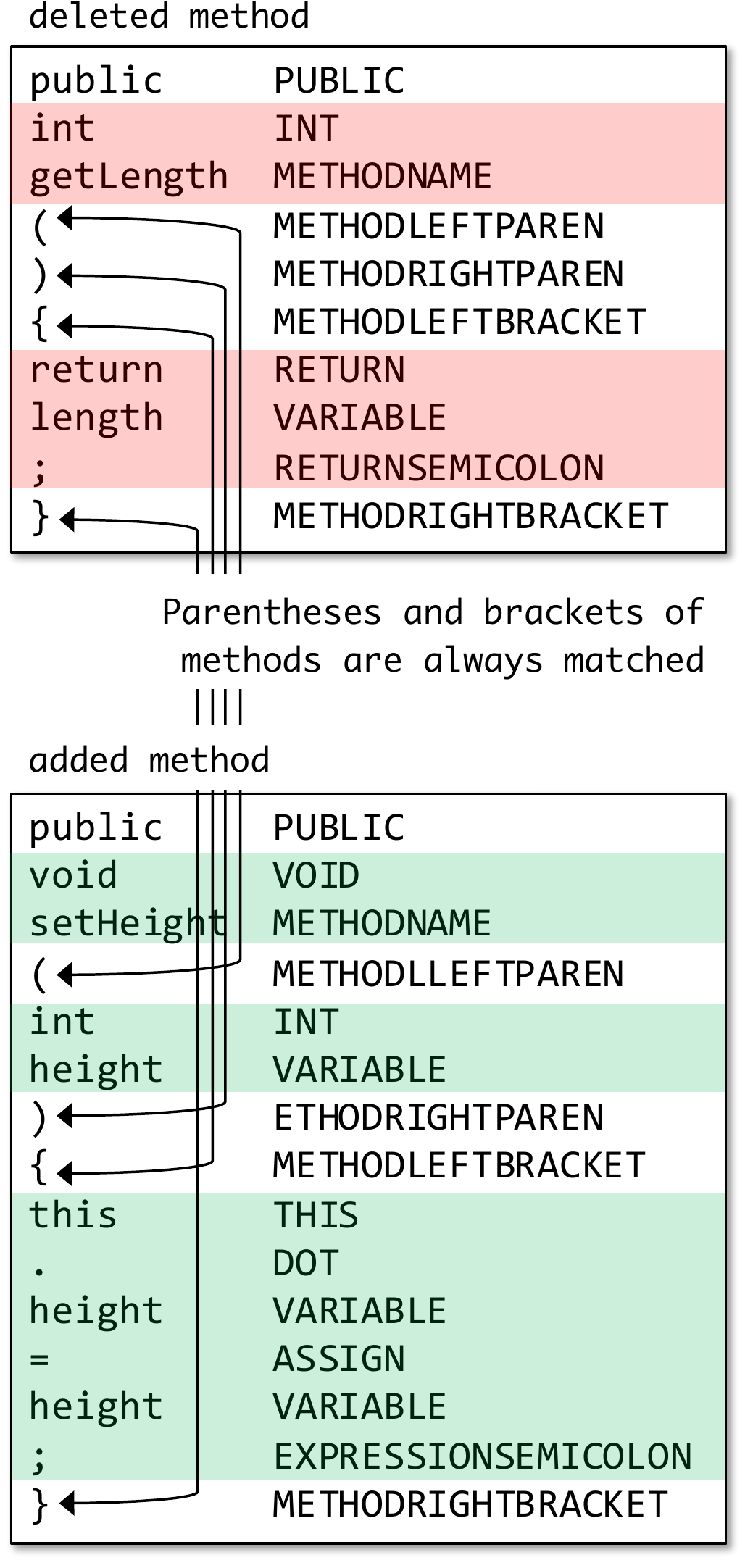}
    \label{fig:Heuristic2Without}
  }
  \subfloat[w/ Heuristic-2.]{
    \includegraphics[bb=0 0 296 623,width=0.23\textwidth]{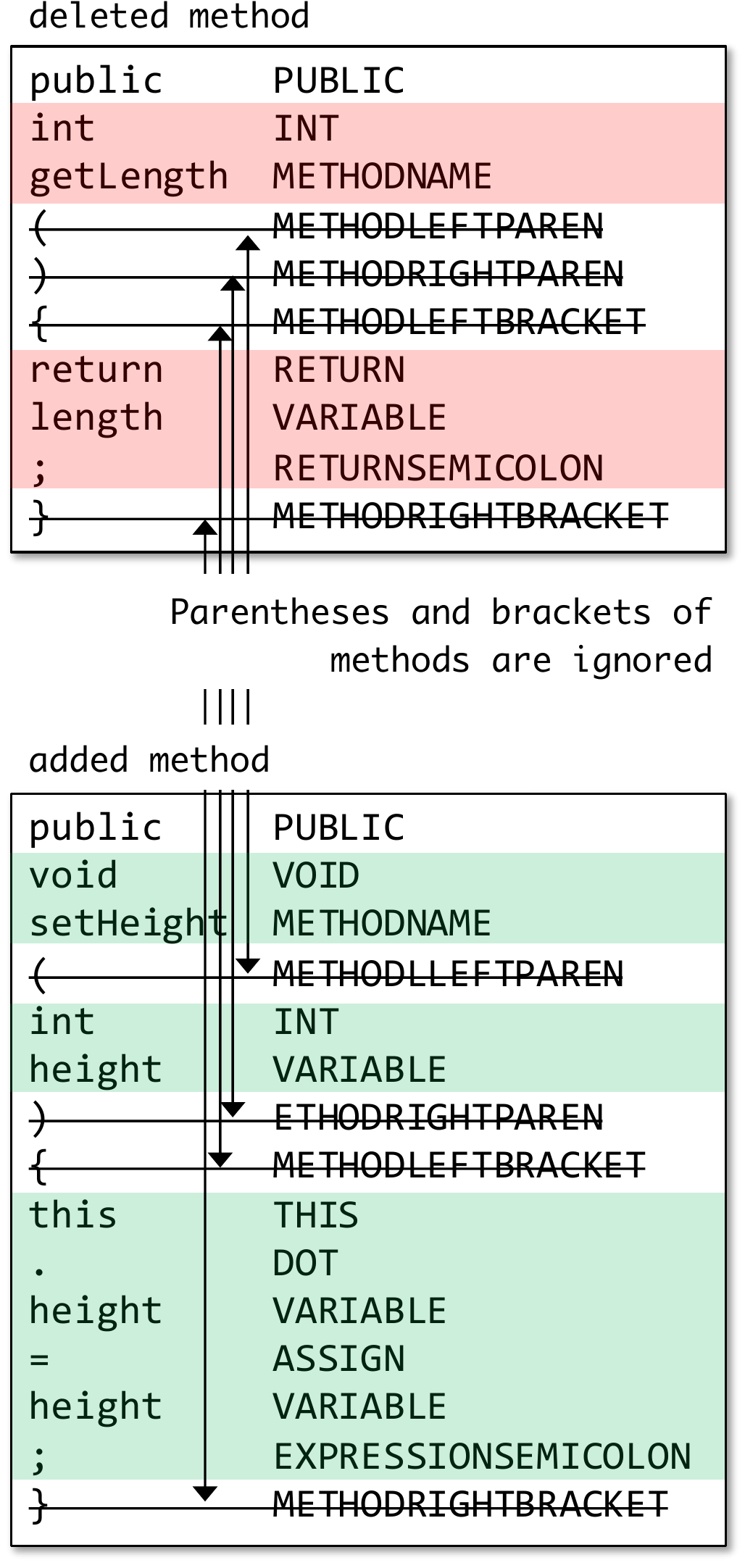}
    \label{fig:Heuristic2With}
  }
  \caption{Tracking files w/ and w/o Heuristic-2.}
  \label{fig:Heuristic2}
  %\vspace{-1.0em}
\end{figure}

%\vspace{1.0em}
%\textbf{Heuristic-2}\\
\subsubsection{Heuristic-2}
The parentheses for parameters and the brackets for method bodies are omnipresent in compilable Java methods.
The fact means that at least the four tokens always match between any Java methods.
Thus, the similarity between non-related methods gets inappropriately higher.
If methods include many tokens, the impact of the four tokens is negligible.
However, if methods are small such as getters and setters, the impact of the four tokens become serious.
Consequently, we decided not to put the four tokens into files for methods.
By removing the four tokens, we prevent the similarity of two non-related methods from getting higher inappropriately.

Figure~\ref{fig:Heuristic2} shows how Heuristic-2 affects tracking.
This example shows a similarity calculation between \texttt{getLength} (before refactoring) and \texttt{setHeight} (after refactoring) in Figure~\ref{fig:TrackingModel}.
A lower similarity between the two methods is better because they are different methods.
In the case that we calculate a similarity without Heuristic-2, the similarity becomes 5/10 (=50\%). 
However, in the case that we adopt Heuristic-2, the similarity becomes 1/6 (=17\%) because the four tokens are ignored.

\section{Implementation}
\label{sec:implementation}

We have implemented a tool based on \FinerGit.
Our tool is open to the public in \GitHub, and anyone can use it freely. 
Our tool takes a \Git repository of a Java project, and it outputs another \Git repository where each Java method gets extracted as a file.
%Each line of the files includes only a single token.
In \FinerGit repositories, method files have extension \texttt{.mjava}.
By executing \texttt{git-log} command with option \texttt{--follow} for \texttt{.mjava} files, we can get their histories.

The name of a method file includes the information of the signature of the method and the class name including the method so that the file name occasionally becomes very long.
Very long file names are not compatible with widely-used operating systems.
For example, in the case of Windows 10, the absolute path of a file must not exceed 260 characters.
If a file name violates the restriction, its file cannot be accessed with Windows' file manager and some other problems occur.
In the case of Linux and MacOS, a file name (not a file path) must not exceed 255 characters.
For practical use in such widely-used operating systems, if a file name becomes longer than the restriction of operating systems, our tool cuts the file name in the middle and then it appends a hash value that is calculated from the entire file name.
This manipulation can shorten the file name \R{\ID{\#2.8}while} keeping its identity.

There are three types of comments in Java source code: line comments, block comments, and Javadoc comments.
Line and block comments are removed from \texttt{.mjava} files while Javadoc comments are included in \texttt{.mjava} files as they are in \texttt{.java} files.
This means that a Javadoc comment exists in the header part of \texttt{.mjava} file if its original method has it.

\ID{\#2.4}
Our tool also has a function to extract each field in Java source code as a single file.
Files for fields have extension \texttt{.fjava}. 
A field declaration includes multiple tokens such as field name, field type, modifiers, initializations, and annotations. 
Thus, fields can be tracked as well as methods by placing a single token on a line.
A file name for a Java field include the following information:
\begin{itemize}
  \item a class including the field,
  \item access modifiers of the field,
  \item a type of the field, and
  \item a name of the field.
\end{itemize}
For example, the file name for field \texttt{length} in Figure~\ref{fig:TrackingModel} becomes as follows.

\vspace{0.5em}
\verb|Person#private_int_length.fjava|
\vspace{0.5em}

Including the above information in the file name reflects code changes around a given field as follows.
\begin{itemize}
  \item If the name of class including the given field is changed, the file name of the given method gets changed, but its contents are not changed.
  \item If another method or field in the class including the given field is changed, neither file name nor the contents of the given method are changed.
  \item If the access modifiers, type, or name of the field is changed, the file name of the given field gets changed and its contents are also changed.
  \item If the annotations and/or initializations of the field are changed, the file name of the given field does not get changed while its contents get changed.
\end{itemize}

In \Historage repository, a file path of a method includes its signature information.
\Historage makes a directory for each Java class.
Methods included in a class are placed in its corresponding directory.
\R{\ID{\#2.3}On the other hand, our technique places files of Java methods in the same directory of their original Java files.}
A reason why \R{\FinerGit does} not make new directories for Java classes is that the conversion time of \Historage is long and making a large number of directories in the conversion process is a factor of taking a long time.
Both \R{\FinerGit} and \Historage make a large number of files because each Java method is extracted as a single file, but our technique does not make new directories for Java classes.
\R{\ID{\#2.3}In both \FinerGit and \Historage, file name collisions for extracted files do not occur as long as their source code is compilable.}

\section{Evaluation}
\label{sec:evaluation}

We evaluated \FinerGit by comparing it with \Historage~\cite{hata2011iwpse}.
We did not use the published version of \Historage implementation\footnote{\url{https://github.com/niyaton/kenja}} but we added \Historage's functionality to our tool.
By using the same implementation for \FinerGit and \Historage, we can avoid different tracking results due to the differences in implementation details.
For example, original \Historage makes directories for each Java class while our \Historage implementation outputs files of Java methods in the same directory as their original files.
The file name convention of our \Historage implementation is the same as \FinerGit.
Thus, in this way, we can evaluate how much method trackability with \Git mechanisms gets improved by \FinerGit.

We selected 182 Java projects in \GitHub as our evaluation targets.
In the process of our target selection, we used Borges dataset~\cite{borges2016icsme}.
This dataset includes 2,279 popular projects in \GitHub.
Firstly, we extracted 202 projects that are labeled as ``Java projects''.
Borges et al.\ classified the projects in the dataset into six categories: \textit{Application software}, \textit{System software}, \textit{Web libraries and frameworks}, \textit{Non-web libraries and frameworks}, \textit{Software tools}, and \textit{Documentation}.
Secondly, we extracted 185 projects that are other than \textit{Documentation} projects because they are repositories with documentation, tutorials, source code examples, etc. (e.g., java-design-patterns\footnote{\url{https://github.com/iluwatar/java-design-patterns}}).
\textit{Documentation} projects are outside of the scope of this evaluation.
Then, we cloned the 185 repositories to our local storage on March 4th 2019.
Unfortunately, we found that three of the 185 projects did not include \texttt{.java} file.
The three projects (google/iosched, afollestad/material-dialogs, and googlesamples/android-topeka) are Kotlin projects.
Finally, we removed the three projects from the 185 projects.

\begin{figure}[t]
  %\vspace{-0.5em}
  \centering
  \includegraphics[bb=0 0 530 436,width=0.40\textwidth]{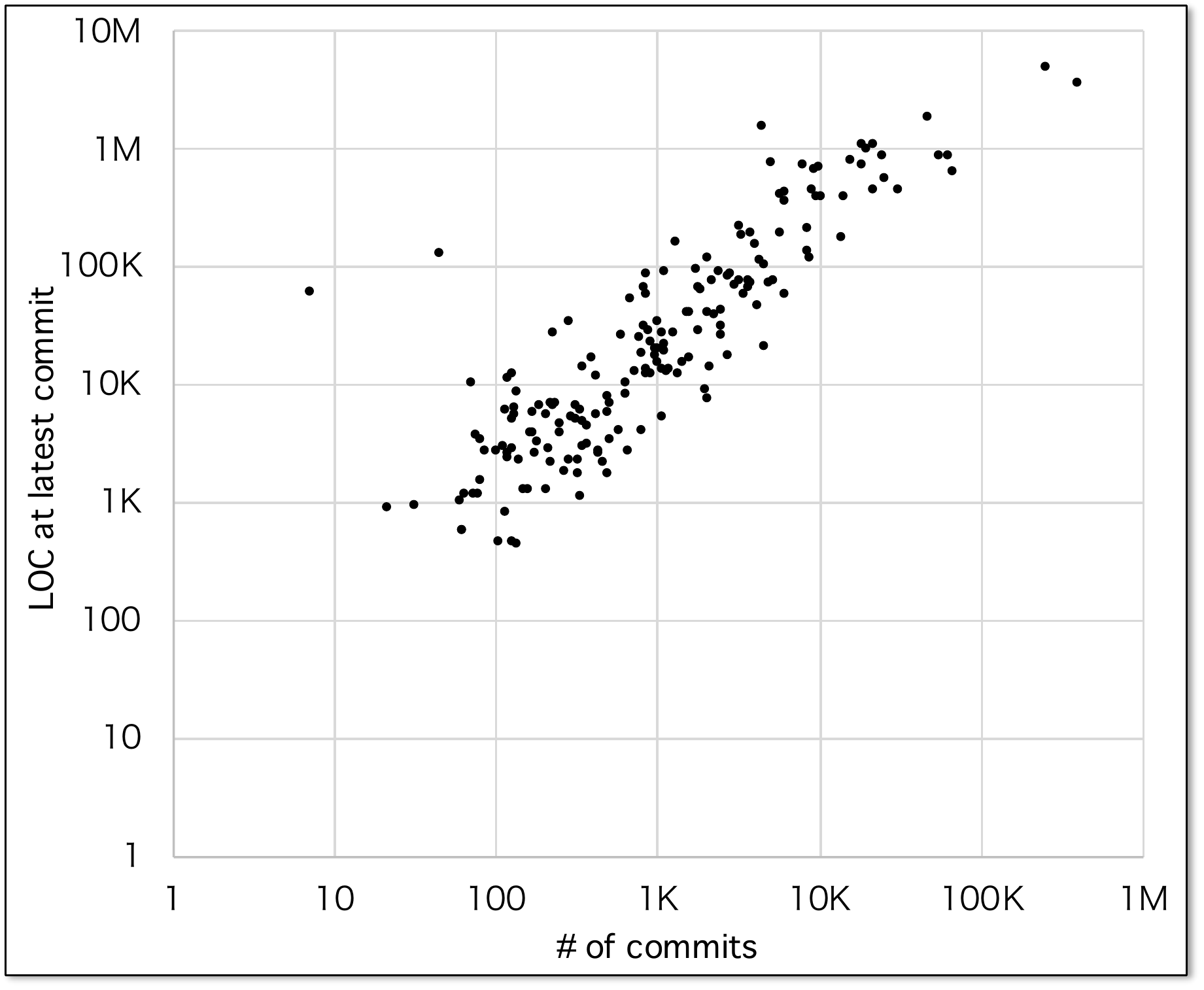}
  \caption{Project size.}
  \label{fig:ProjectSize}
  %\vspace{-1.0em}
\end{figure}

Figure~\ref{fig:ProjectSize} shows the distributions of the number of commits and LOC of the target projects.
The two largest repositories in the targets are platform\_frameworks\_base\footnote{\url{https://github.com/aosp-mirror/platform_frameworks_base}} and intellij-community\footnote{\url{https://github.com/JetBrains/intellij-community}}.
The two repositories include approximate 380K and 240K commits, and their latest revisions consist of about 3.7M and 5.0M LOC, respectively.

We generated \FinerGit repositories and \Historage ones from the 182 target projects.
Herein, \FinerGit repositories have the file format of including a single token per line with the two heuristics while \Historage repositories have the same line format as the original repositories.

We have evaluated \FinerGit from the five viewpoints:
\begin{itemize}
  \item tracking accuracy,
  \item heuristics impacts,
  \item project-level tracking results,
  \item method-size-level tracking results, and
  \item execution time.
\end{itemize}

Hereafter in this section, we report the results in detail.

\subsection{Tracking Accuracy}
\label{sec:evaluation:accurary}

It is not realistic to manually check whether \FinerGit generates correct tracking results for each method in the target projects.
Thus, we make an oracle for a method for each target project with the following procedure.
\begin{enumerate}
  \item A method was randomly selected from each target project. In total, 182 methods were selected.
  \item Each of the methods in \FinerGit repositories was tracked with the following command.
  %\vspace{0.5em}
  %\begin{verbatim}

  \noindent
  \verb!  > git log --follow -U15 -M20% -C20% -p!\\
  \verb!      -- !\textit{path/to/method}\verb!.mjava!
  %\end{verbatim}
  %\vspace{-0.5em}

  \noindent
  With the above command, a specified file is tracked even if the file was renamed.
  If there is a file that has a 20\% or more similarity, \Git regards that file renaming or copying occurred.
  \item The tracking results were examined, and oracles of renaming and copying history were made by two of the authors independently.
  Each author spent several hours on this task.
  The two authors made different oracles for 34 out of the 182 methods.
  \item The two authors discussed the 34 methods so that they obtain consensus for them.
  After a two-hour discussion, they got consensus oracles for the 34 methods.
\end{enumerate}
With the above procedure, we obtained consensus oracles of tracking results for the 182 methods.
Finally, we obtained the resulting oracle set consisting of 426 renaming/copying changes for the 182 methods in total.

Next, we track the methods in \FinerGit's repositories and \Historage's ones with different thresholds.
We used the following command to count how many times \Git found renaming and copying with a specified threshold.

%\begin{verbatim}
  \noindent
  \verb!  > git log --follow --oneline -M!\textit{t}\verb! -C!\textit{t}\verb! -p!\\
  \verb!      -- !\textit{path/to/method}\verb!.mjava!\\
  \verb!      | grep -e "^rename from\|^copy from"!\\
  \verb!      | wc -l!
%\end{verbatim}

\noindent
In the above command, \textit{t} is the threshold that \Git regards given two files have a renaming or copying relationship.
We tracked the target methods with 13 different thresholds (i.e., 20\%, 25\%, 30\%, $\ldots$, 80\%).
If tracking results for a method include a higher number of renaming/copying than its oracle, we regard renaming/copying in the over-tracking part as false positives.
If tracking results for a method include a lower number of renaming/copying than its oracle, we regard renaming/copying that are not detected as false negatives.
We calculated precision, recall, and F-measure for each threshold by summing up the number of false positives and false negatives of all the methods.

\begin{figure}[t]
  %\vspace{-1.0em}
  \centering
  \includegraphics[bb=0 0 578 309,width=0.48\textwidth]{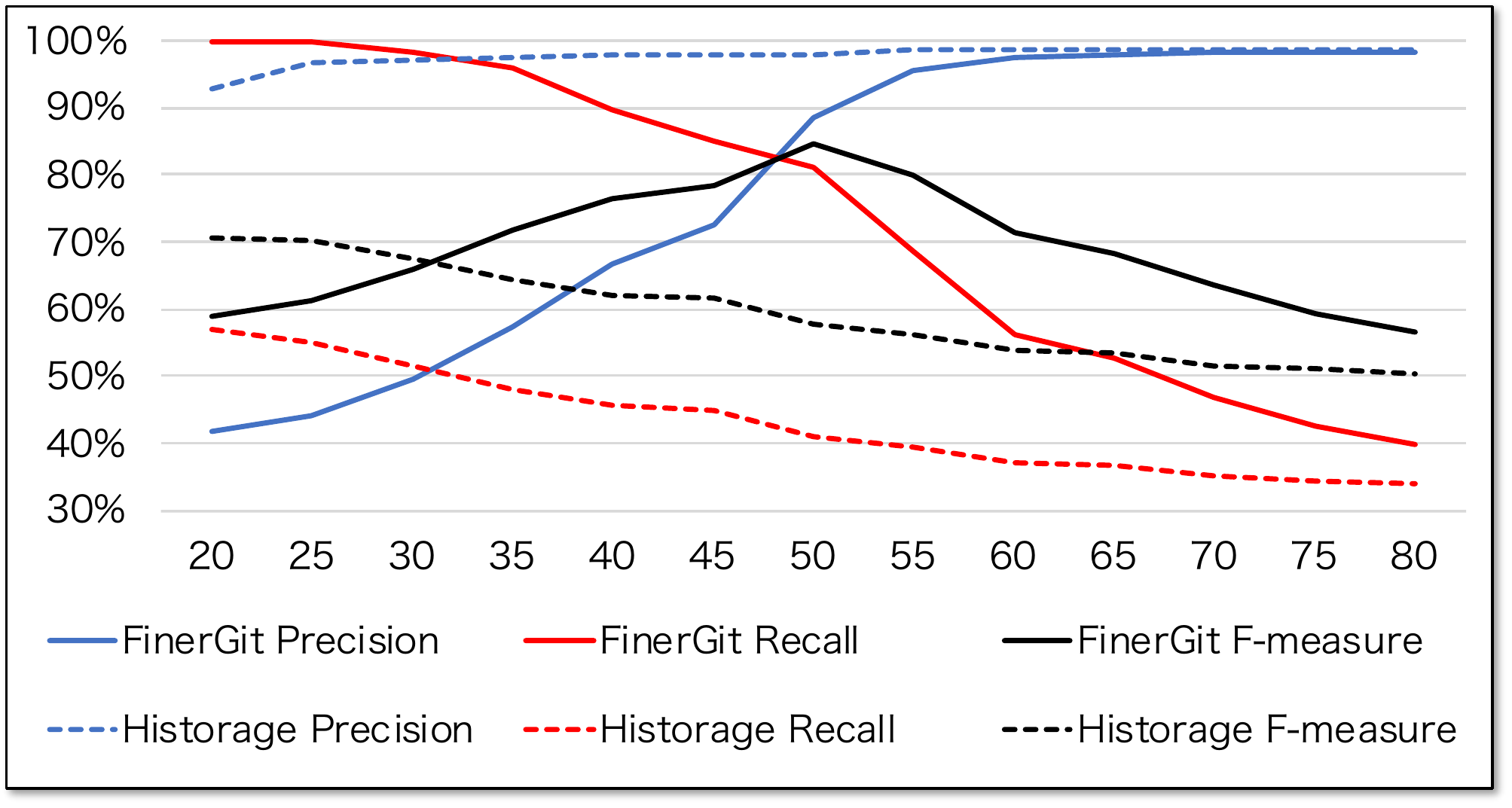}
  \caption{Precision, recall, and F-measure values.}
  \label{fig:PrecisionRecall}
  %\vspace{-1.0em}
\end{figure}

Figure~\ref{fig:PrecisionRecall} shows how precision, recall, and F-measure changes according to given thresholds.
The graphs of \Historage and \FinerGit have the following features.
\begin{itemize}
  \item Precision of \Historage is very high. \Historage has 93.01\% of precision even in the case of threshold 20\%.
  \item Recall of \Historage is low. \Historage has only 57.04\% of recall in the case of threshold 20\%.
  \item \FinerGit has high precision in the case of high thresholds, but precision gets rapidly decreased for lower thresholds.
  \item \FinerGit has higher recall than \Historage for all the thresholds. The recall differences between \FinerGit and \Historage get bigger for lower thresholds.
\end{itemize}
\Historage has a low possibility to track wrong methods while it often misses renaming and copying.
On the other hand, in \FinerGit repositories, precision gets decreased for lower thresholds while recall improves much.
The highest F-measure on \FinerGit is 84.52\% on threshold 50\% while the highest F-measure on \Historage is 70.72\% and 70.23\% on thresholds 20\% and 25\%, respectively.

\subsection{Heuristics Impacts}

\begin{figure}[t]
 %\vspace{-1.0em}
  \centering
  \subfloat[Precision.]{
   \includegraphics[bb=0 0 458 249,width=0.42\textwidth]{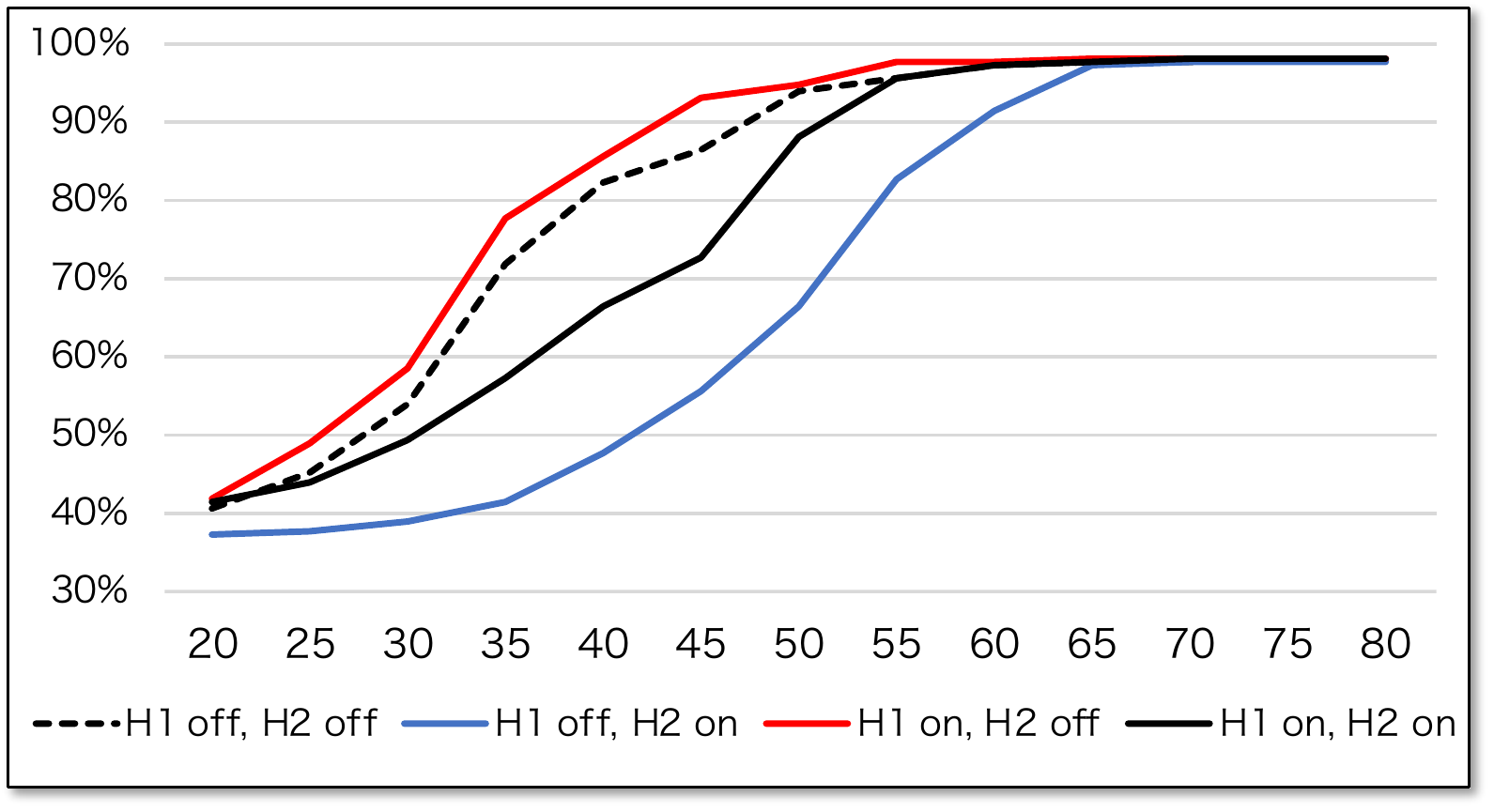}
   \label{fig:HeuristicsPrecision}
  }\\
  \subfloat[Recall.]{
    \includegraphics[bb=0 0 459 249,width=0.42\textwidth]{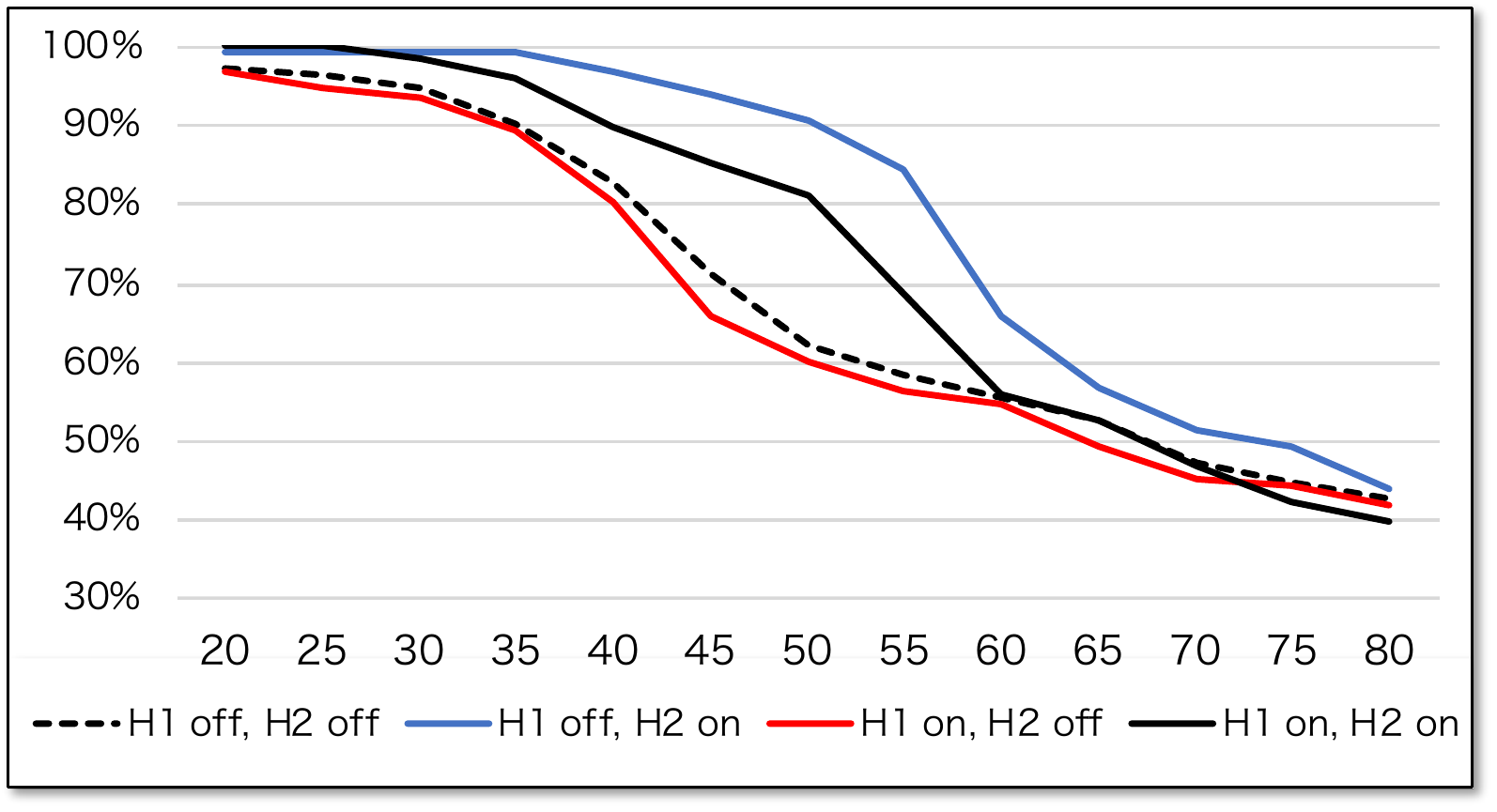}
    \label{fig:HeuristicsRecall}
  }\\
  \subfloat[F-measure.]{
    \includegraphics[bb=0 0 459 249,width=0.42\textwidth]{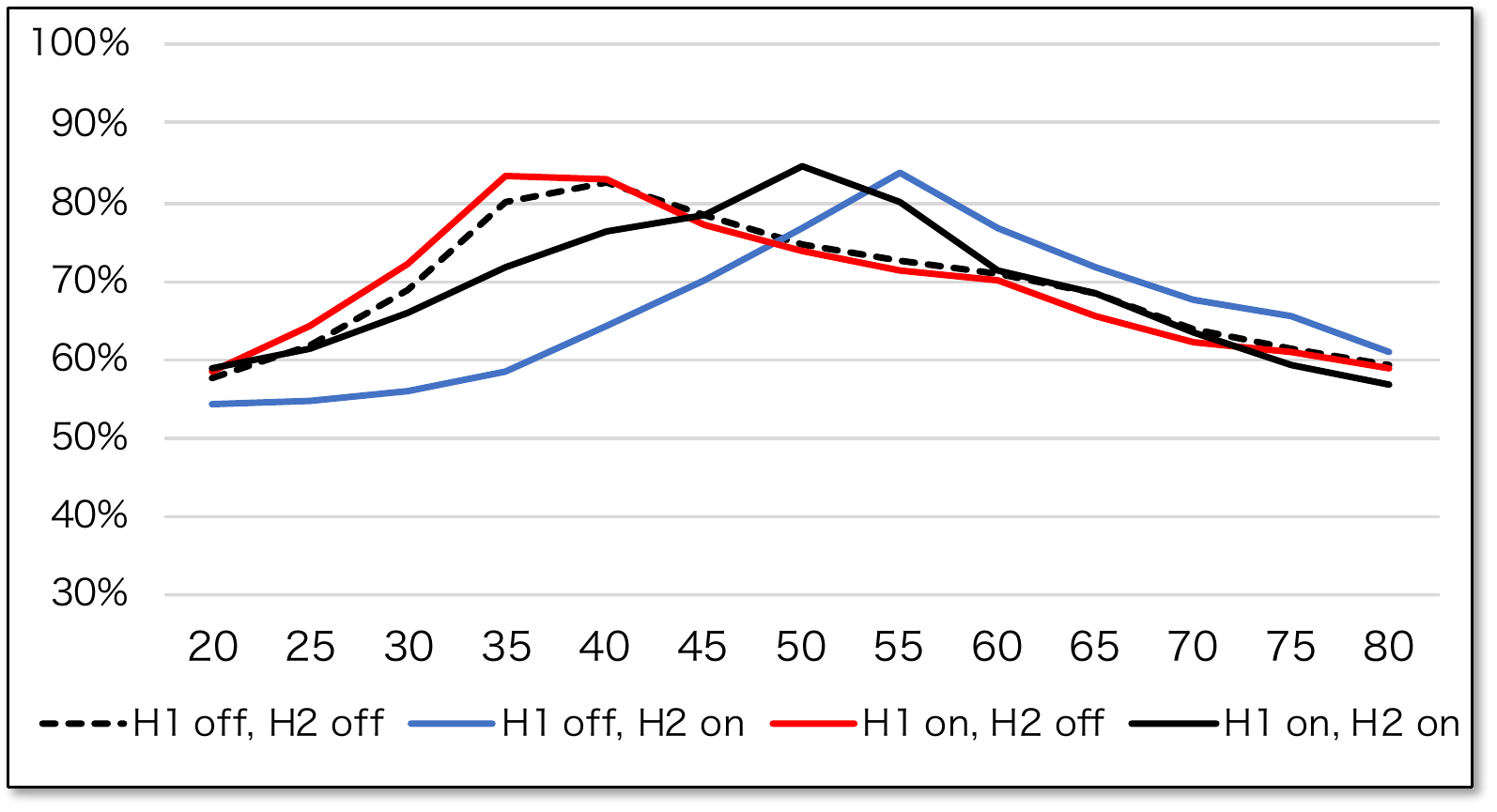}
    \label{fig:HeuristicsFmeasure}
  }\\
  \subfloat[Rename count.]{
    \includegraphics[bb=0 0 459 249,width=0.42\textwidth]{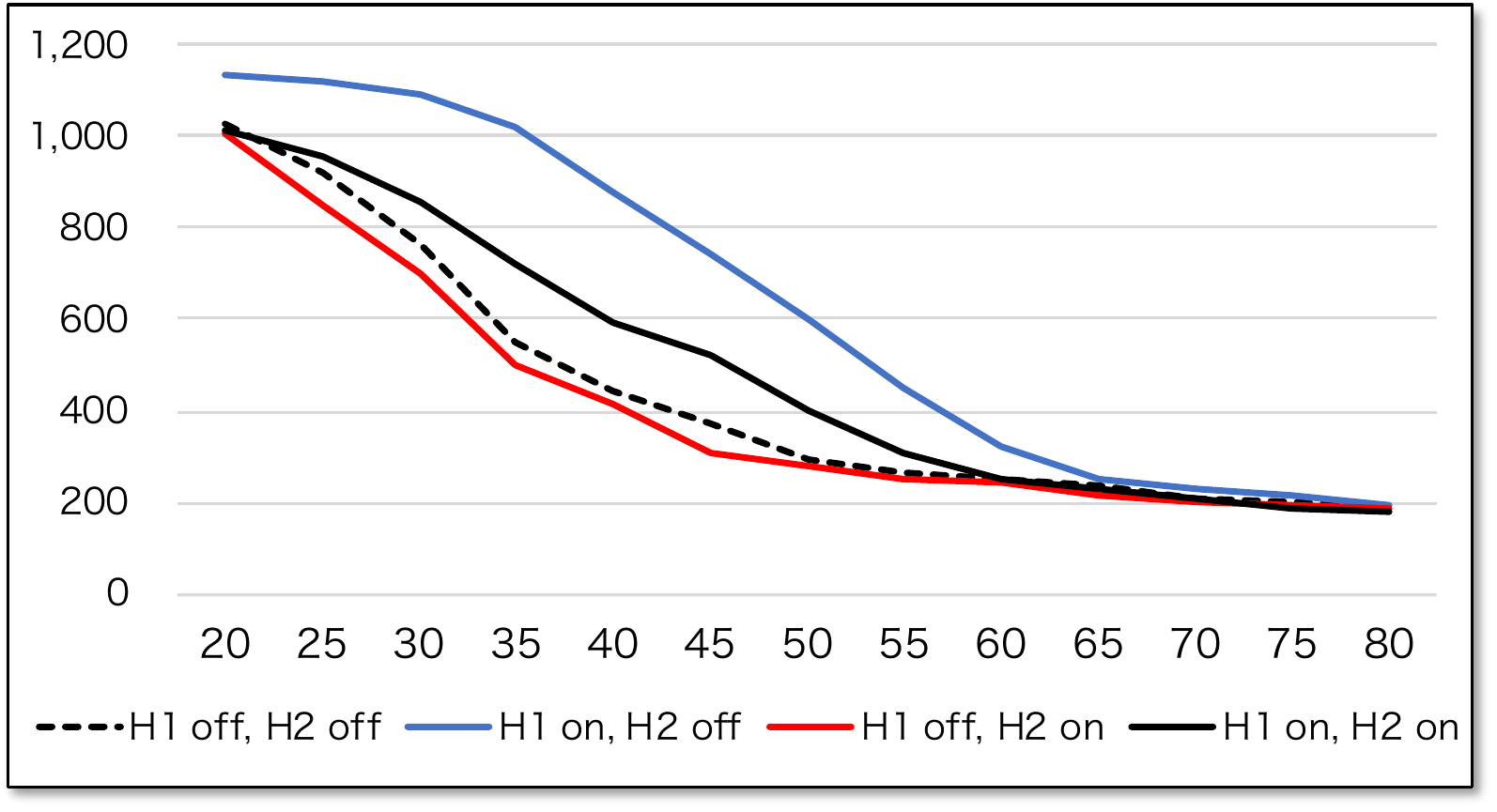}
    \label{fig:HeuristicsRenameCount}
  }
  \caption{Precision, recall, F-measure, and rename count when heuristics 1 and 2 are on and/or off.}
  \label{fig:HeuristicsImpacts}
  %\vspace{-1.0em}
\end{figure}

To reveal how each heuristic impacts on method tracking, we measured precision, recall, and F-measure and we also counted found renames for the following four types of fine-grained repositories.
The target methods are the same as Subsection~\ref{sec:evaluation:accurary}.
Herein, rename count means the sum of found renames for all the target methods in a type of repositories.
\begin{description}
 \item[H1 OFF, H2 OFF:]neither heuristics are applied to.
 \item[H1 ON, H2 OFF:]only Heuristic-1 is applied to.
 \item[H1 OFF, H2 ON:]only Heuristic-2 is applied to.
 \item[H1 ON, H2 ON:]both heuristics are applied to. This is the same repository as what we used in Subsection~\ref{sec:evaluation:accurary}.
\end{description}
Figure \ref{fig:HeuristicsImpacts} shows the results.
Applying only Heuristic-1 makes it possible to find more renaming so that precision gets decreased while recall gets increased.
On the other hand, applying only Heuristic-2 slightly shorten method tracking.
As a result, precision gets increased while recall gets decreased.
The reasons why applying Heuristic-1 and Heuristic-2 have opposite impacts on method tracking are as follows.
\begin{itemize}
  \item Applying Heuristic-1 reduces similarities between methods. How much the similarities are decreased depends on the contents on methods. Thus, a different method can be tracked at a commit compared to the case that Heuristic-1 is not applied to.
  \item Applying Heuristic-2 reduces similarities between all methods. Unlike Heuristic-1, Heuristic-2 does not make a different method tracked. Thus, Heuristic-2 just shortens method tracking.
\end{itemize}

Table~\ref{tbl:maxFmeasure} shows the maximum F-measure for each type of finer-grained repositories.
In this table, the maximum F-measure is the greatest F-measure in all data.
All types have almost the same maximum values.
This table also shows the maximum recall when we track methods with over 95\% precision.
These results show that more method renames are found with keeping 95\% precision by applying both heuristics.

\begin{table}[b]
  \centering
  \caption{Maximum F-measure and Maximum Recall} % The maximum F-measure is the greatest F-measure in all data. The maximum recall is the maximum value in case that precision is over 95\%.}
  \label{tbl:maxFmeasure}
  {\small\tabcolsep=0.3em\begin{tabular}{llrr} \hline
    \multicolumn{2}{c}{Repository type} & Max F-measure (thr.) & Max Recall (thr.) \\\hline
    H1 OFF, & H2 OFF & 82.63\% (40\%) & 58.45\% (55\%) \\
    H1 ON, & H2 OFF & 83.77\% (55\%) & 56.81\% (65\%) \\
    H1 OFF, & H2 ON & 83.26\% (35\%) & 60.09\% (50\%) \\
    H1 ON, & H2 ON & 84.52\% (50\%) & 68.78\% (55\%) \\\hline
  \end{tabular}}
\end{table}

\subsection{Project-Level Tracking Results}

\begin{figure}[t]
  %\vspace{-2.0em}
  \centering
  \subfloat[Ratio of different tracking results.]{
   \includegraphics[bb=0 0 234 309,width=0.23\textwidth]{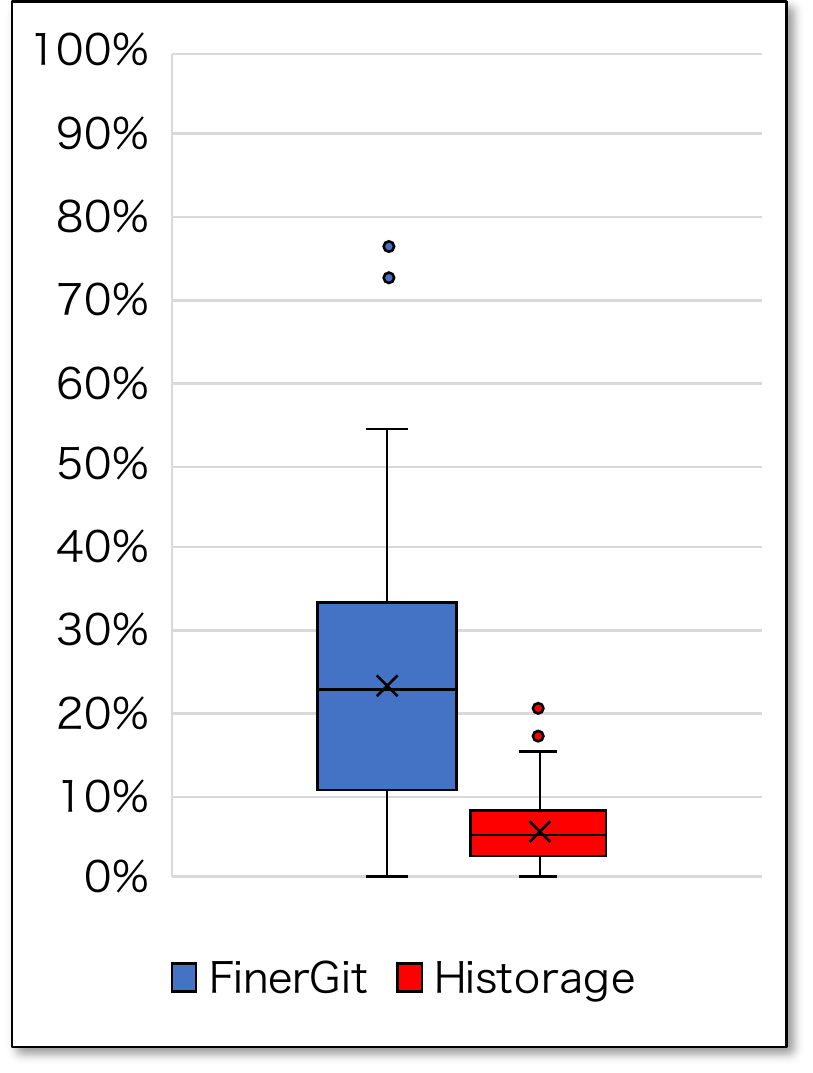}
   \label{fig:ProjectLevelComparison:Ratio}
  }
  \subfloat[Average change counts.]{
    \includegraphics[bb=0 0 232 308,width=0.23\textwidth]{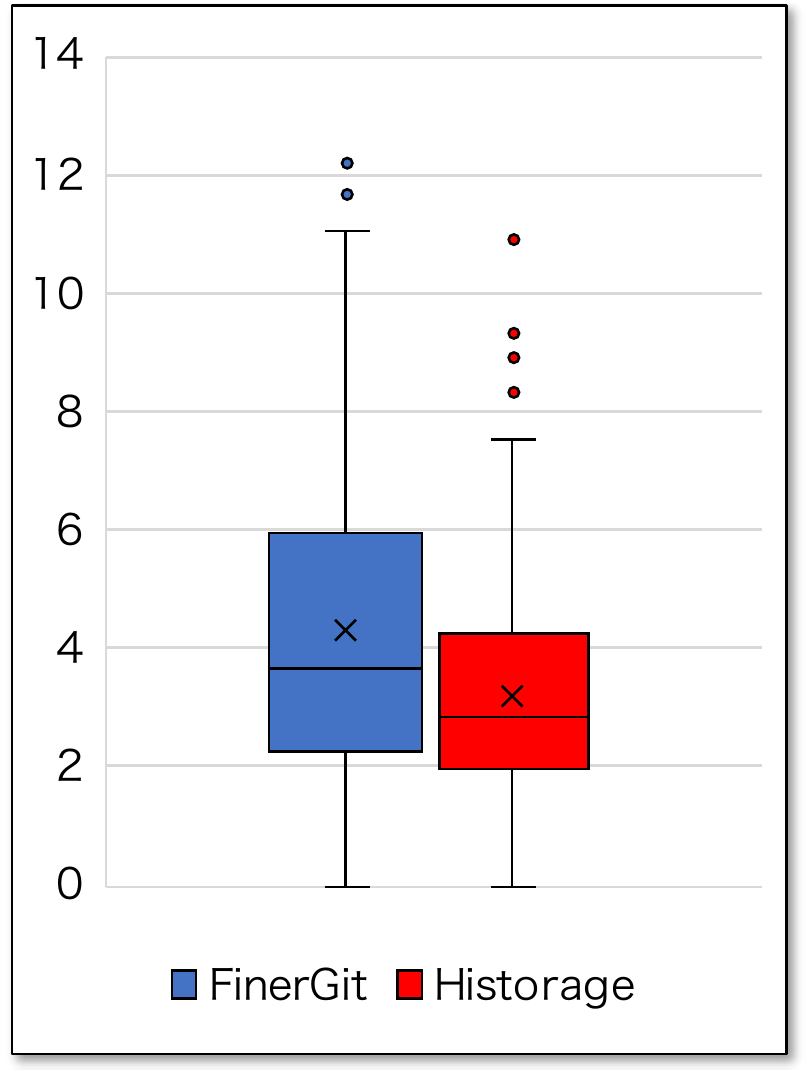}
    \label{fig:ProjectLevelComparison:Count}
  }
  \caption{Project-level comparisons. (a) shows the ratio of methods whose tracking results are different between \FinerGit and \Historage for each project. (b) shows the average of change counts for all the methods for each project.}
  \label{fig:ProjectLevelComparison}
  %\vspace{-1.0em}
\end{figure}

In this evaluation, we measured the ratio of methods whose tracking results are different between the two tools for each project.
We compare how much the number of detected renames is different from \FinerGit and \Historage under the same precision.
As shown in the previous subsection, the two tools have different precision values for different thresholds.
To realize a fair comparison, we decided to select different thresholds for \FinerGit and \Historage that satisfy the following condition: method tracking results with the thresholds have the same precision values and the precision values are as high as possible.
Thus, we used threshold 55\% for \FinerGit and 25\% for \Historage.
The precision of \FinerGit on threshold 55\% is 95.73\%, and \Historage on threshold 25\% is 96.60\%.
Those precision values are almost the same and high enough.

Figure~\ref{fig:ProjectLevelComparison} shows the comparison results.
In Figure~\ref{fig:ProjectLevelComparison}\subref{fig:ProjectLevelComparison:Ratio}, the blue boxplot shows the ratio of methods for which \FinerGit found more renames than \Historage per project and the red boxplot shows the opposite one.
\FinerGit found more renames for 22.71\% methods on average while the ratio of methods that \Historage found more renames than \FinerGit is only 5.26\%.
In Figure~\ref{fig:ProjectLevelComparison}\subref{fig:ProjectLevelComparison:Count}, the blue boxplot shows the average number of changes identified by \FinerGit for all methods of each project.
The red one shows the average number of changes identified by \Historage.
\R{\ID{\#2.8}The median values of those boxplots are 3.67 and 2.86, respectively.
These results mean that \FinerGit can find more renames for all the methods on average.}

Next, we show that the tracking improvement by \FinerGit is effective \R{\ID{\#2.8}via} the following two ways:
\begin{itemize}
 \item considering the fact that some methods were never changed \R{\ID{\#2.8}after their initial creation}, and
 \item conducting statistical testing for the tracking results.
\end{itemize}

% \begin{figure}[t]
%   \centering
%   \includegraphics[width=0.48\textwidth]{UnchangedMethods-crop.pdf}
%   \caption{Ratio of methods that were never changed}
%   \label{fig:UnChangedMethods}
%   \vspace{-1.0em}
% \end{figure}

\subsubsection{Considering Never-Changed Methods}
In software development, some methods are never changed after their \R{\ID{\#2.8} initial creation}.
If the 182 target projects include many never-changed methods, it is quite natural that the comparison results between \FinerGit and \Historage are not so different from each other.
Thus, we investigate how many never-changed methods are included in the projects.
It is not realistic to manually collect real never-changed methods.
In this experiment, we decided to regard methods that both \FinerGit and \Historage were not able to detect any changes as never-changed methods.

% Figure~\ref{fig:UnChangedMethods} shows the ratio of never-changed methods against all the methods per project.
% The 25 percentile, the median, and the 75 percentile are 6.88\%, 15.27\%, and 26.50\%, respectively.
% The fact means that most projects include a non-negligible amount of never-changed methods.
% The scatter diagram of Figure~\ref{fig:UnChangedMethods} shows the relationship between the ratio of never-changed methods and the number of commits in the target projects.
% The scatter diagram tells us that the ratio of never-changed methods is almost the same even though projects include a different number of commits.
% There are three projects where all methods were never changed (\textsf{MaterialDesignLibrary}, \textsf{hackpad}, and \textsf{anthelion}).
% Those repositories include 74, 43, and 7 commits and there are 199, 5,302, and 2,196 methods at the latest commit.
% As a matter of course, there is no tracking difference between \FinerGit and \Historage for the three projects.

\begin{figure}[t]
  \centering
  \includegraphics[bb=0 0 458 272,width=0.45\textwidth]{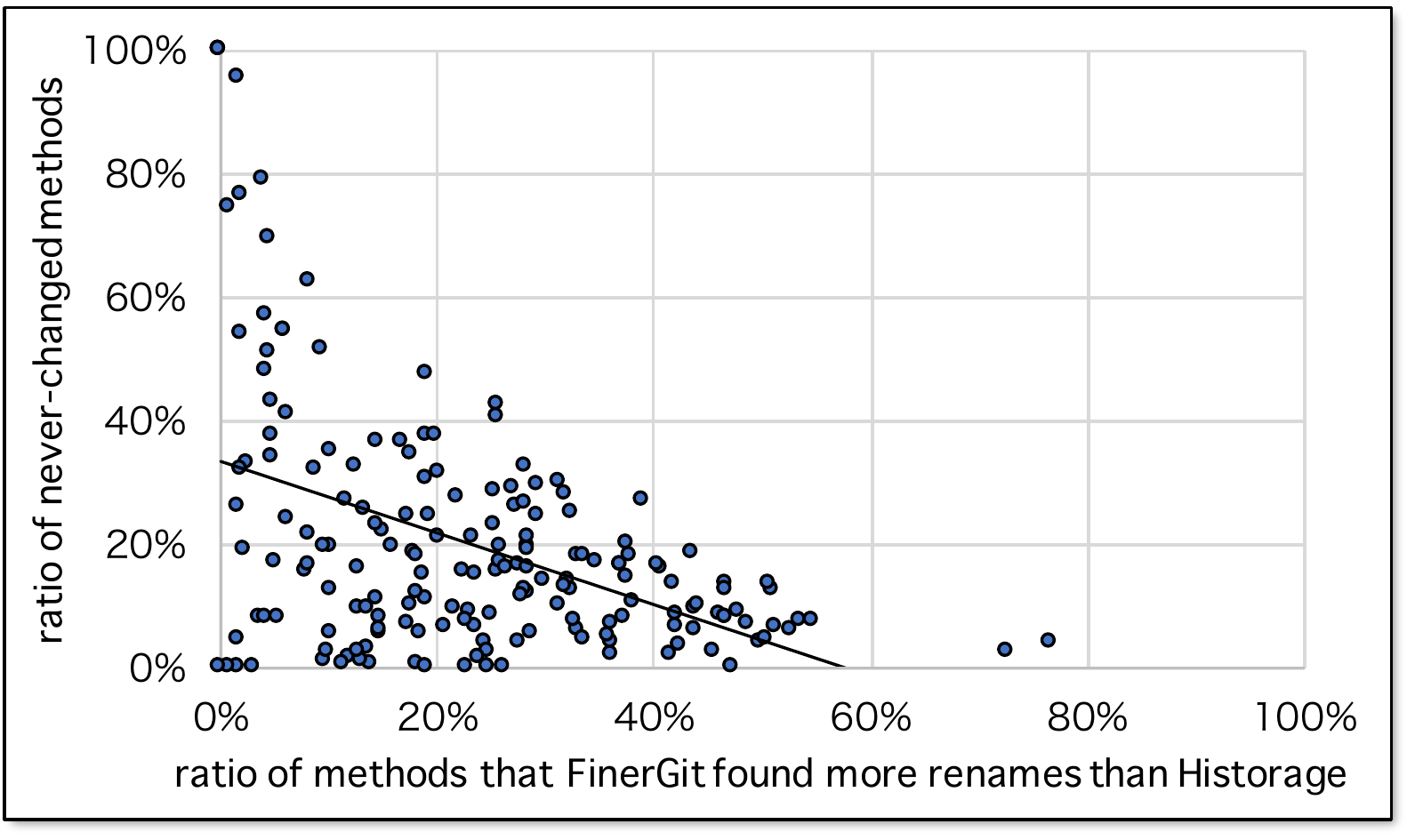}
  \caption{Relationships between the ratio of methods for which \FinerGit found more renames than \Historage and the ratio of never-changed methods.}
  \label{fig:LongerAndChangedRatio}
  %\vspace{-1.0em}
\end{figure}

\begin{figure}[t]
  %\vspace{-1.0em}
  \centering
  \subfloat[w/ never-changed methods.]{
   \includegraphics[bb=0 0 235 328,width=0.23\textwidth]{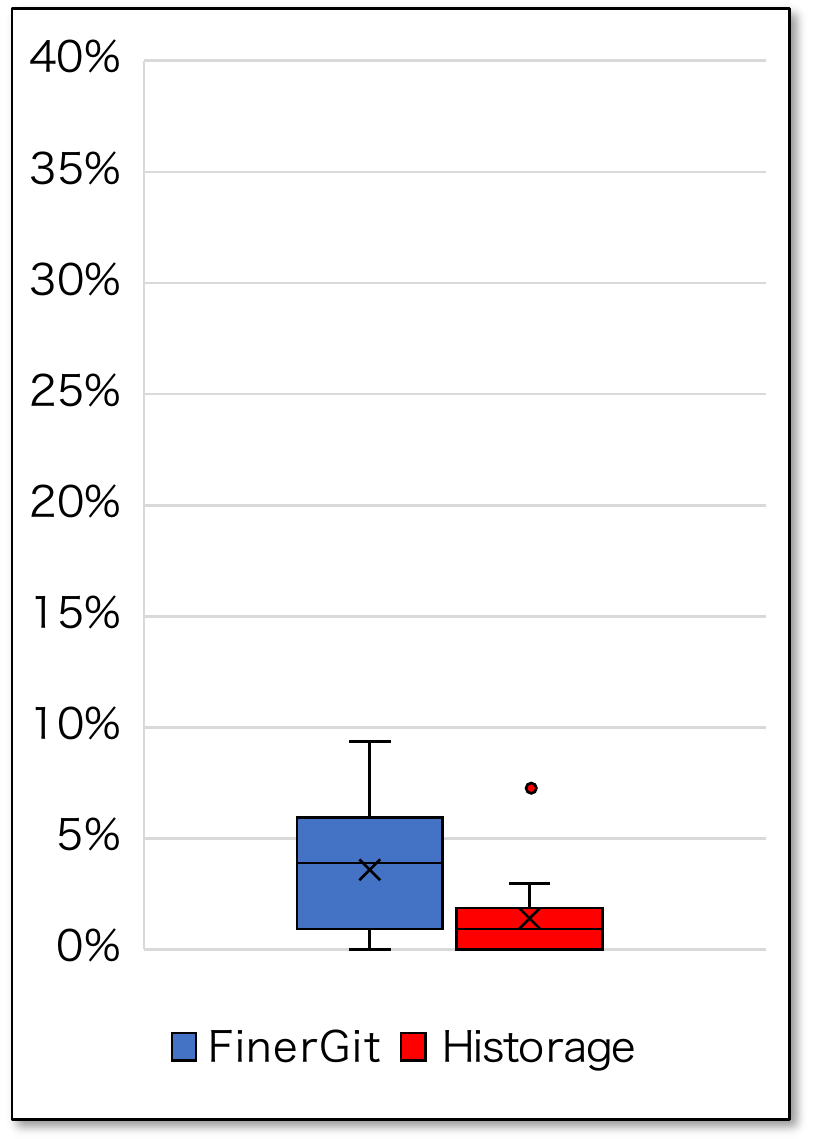}
   \label{fig:ProjectLevel50on}
  }
  \subfloat[w/o never-changed methods.]{
    \includegraphics[bb=0 0 236 330,width=0.23\textwidth]{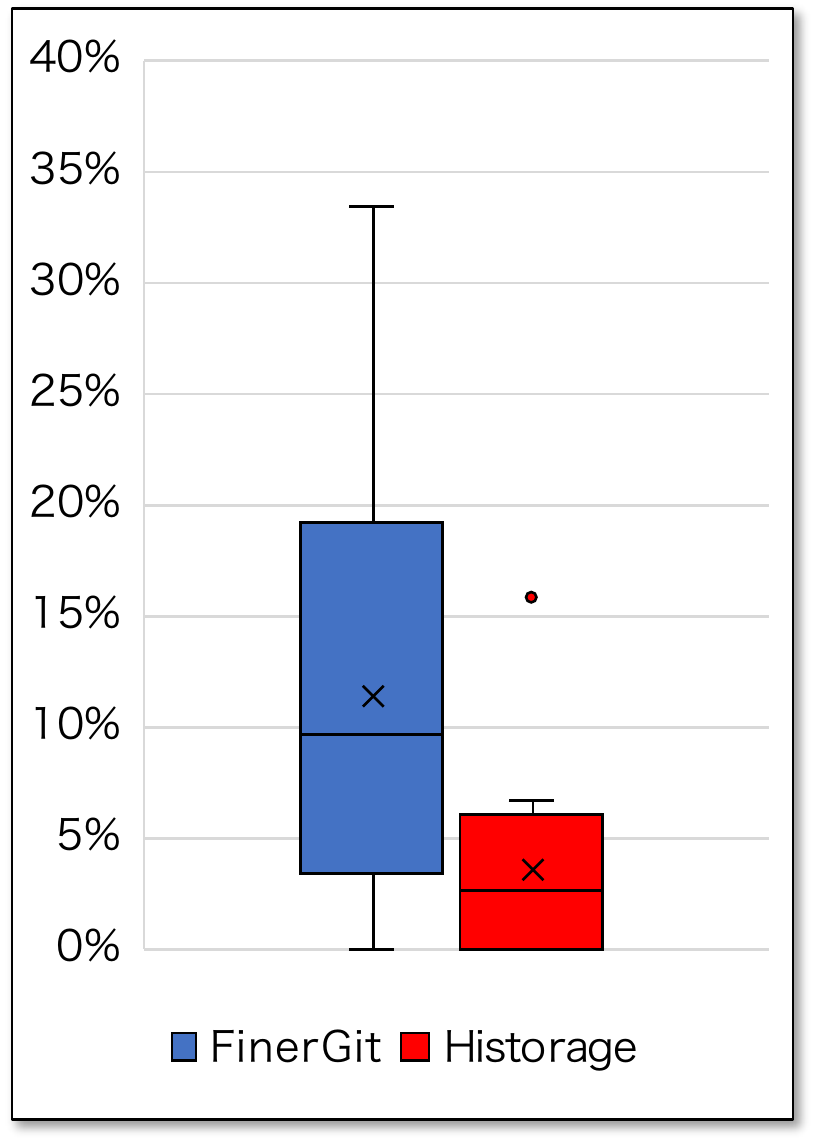}
    \label{fig:ProjectLevel50off}
  }
  \caption{The ratio of methods whose tracking results are different between \FinerGit and \Historage for projects where 50\% or more methods are never-changed ones.}
  \label{fig:ProjectLevel50}
  %\vspace{-1.0em}
\end{figure}

Figure~\ref{fig:LongerAndChangedRatio} shows the relationship between the ratio of never-changed methods and the ratio of methods for which \FinerGit found more renames than \Historage.
The 25 percentile, the median, and the 75 percentile of never-changed methods are 6.88\%, 15.27\%, and 26.50\%, respectively.
The figure indicates that the more never-changed methods there are, the fewer methods \FinerGit found more renames for.
Figure~\ref{fig:ProjectLevel50} shows the same figures as Figure~\ref{fig:ProjectLevelComparison}\subref{fig:ProjectLevelComparison:Ratio} only for the projects that include 50\% or more never-changed methods.
As shown in Figure~\ref{fig:ProjectLevel50}\subref{fig:ProjectLevel50on}, the differences between \FinerGit and \Historage are small because the majority of their methods is never-changed.
Figure~\ref{fig:ProjectLevel50}\subref{fig:ProjectLevel50off} shows the differences after we removed never-changed methods from the projects.
We can see that the differences between the two tools get much larger.
MSR approaches are naturally applied to methods that have change histories.
Never-changed methods are exempt from MSR approaches.

We also investigated how many methods only \FinerGit or \Historage found at least a change for.
The former number is 97,629 and the latter one is 35,553.
They are 5.52\% and 2.01\% of all methods, respectively.
Finding changes for more methods means that various MSR approaches requiring past changes can be applied more broadly.

\subsubsection{Conducting Statistical Testing}
We applied Paired Wilcoxson's signed ranked test to the comparison results between \FinerGit and \Historage shown in Figure~\ref{fig:ProjectLevelComparison}.
The test showed that the comparison results include significant differences regarding both aspects of the ratio ($p$-value $<$ 0.001) and average change counts ($p$-value $<$ 0.001).
We also applied Cliff's Delta to the comparison results to see the effect size.
The resulting values were computed as 0.712 for the ratio and 0.221 for the average change counts, which revealed a \emph{large} and a \emph{small} effect size of the improvement achieved by using \FinerGit, respectively.
%In Cliff's Delta, if the value is 0.474 or higher, the differences are regarded as large.
Consequently, we can say that \FinerGit significantly improves tracking Java methods compared to \Historage.

\subsection{Method-Size-Level Tracking Results}

\begin{figure}[t]
  \centering
  \subfloat[Ratio of methods for which \FinerGit or \Historage found more renames than the other tool.]{
   \includegraphics[bb=0 0 684 227,width=0.475\textwidth]{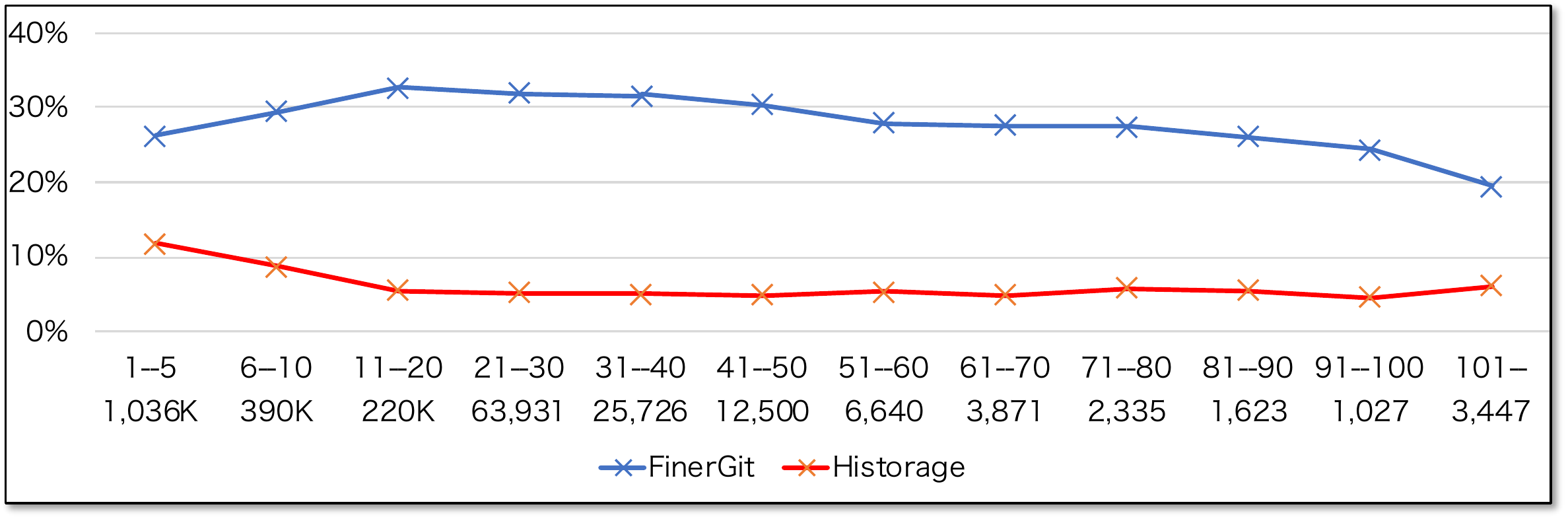}
   \label{fig:MethodSizeDiffereneRatio}
  }\\
  \subfloat[Average renames that were found by \FinerGit or \Historage.]{
    \includegraphics[bb=0 0 684 229,width=0.475\textwidth]{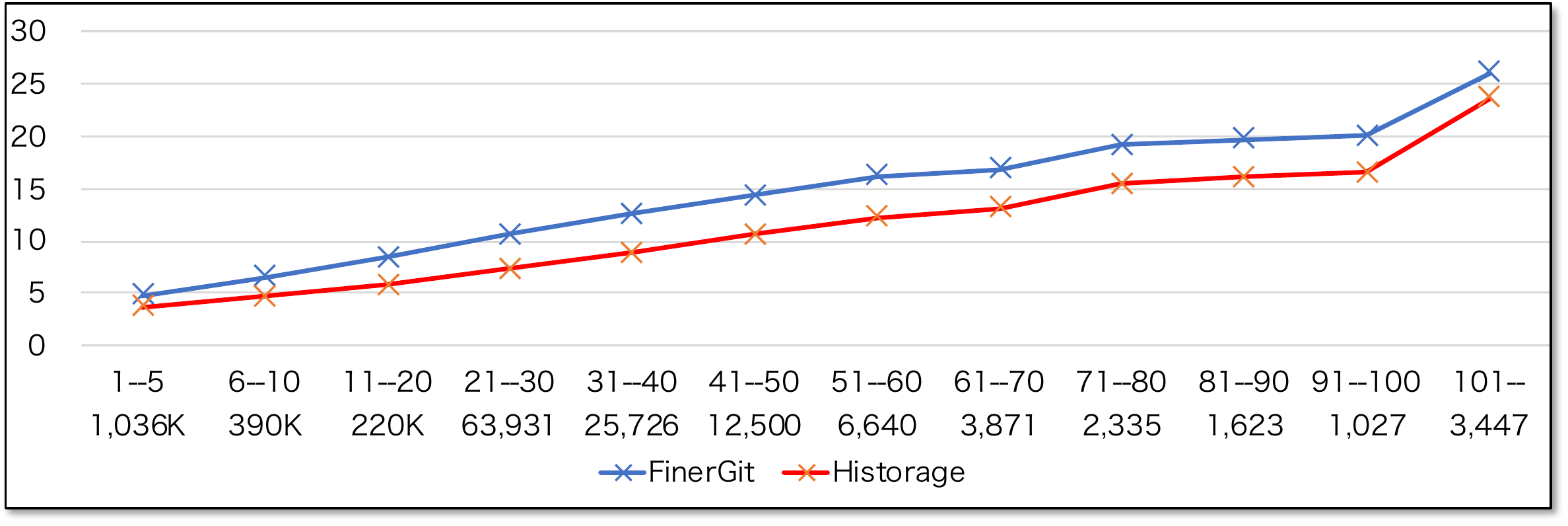}
    \label{fig:MethodSizeRenameCount}
  }
  \caption{Comparison based on method size.}
  \label{fig:MethodSizeComparisonResults}
  %\vspace{-1.0em}
\end{figure}

We also conducted comparisons based on method size.
In this comparison, we made several method groups based on their size.
Then, we compared the tracking results for each group.
Figure~\ref{fig:MethodSizeComparisonResults} shows the comparison results.
We can see that there are 1,036K methods whose LOC is in the range between 1 and 5.
Herein, the LOC was computed using the original format, not the single-token-per-line one.
\FinerGit generated longer tracking results for 26.21\% of the 1,036K methods.
Our research motivation was improving the trackability for small methods, but surprisingly \FinerGit improved the trackability for methods of any size.

This figure also shows the average rename counts that were found by \FinerGit and \Historage.
We can see that \FinerGit found more renames for methods of any size than \Historage.
Interestingly, more renames tend to be found for larger methods by both tools.

Consequently, we conclude that the method tracking capability of \FinerGit is higher than \Historage.

\subsection{Execution Time}

\begin{figure}[t]
  \centering
  \includegraphics[bb=0 0 534 329,width=0.45\textwidth]{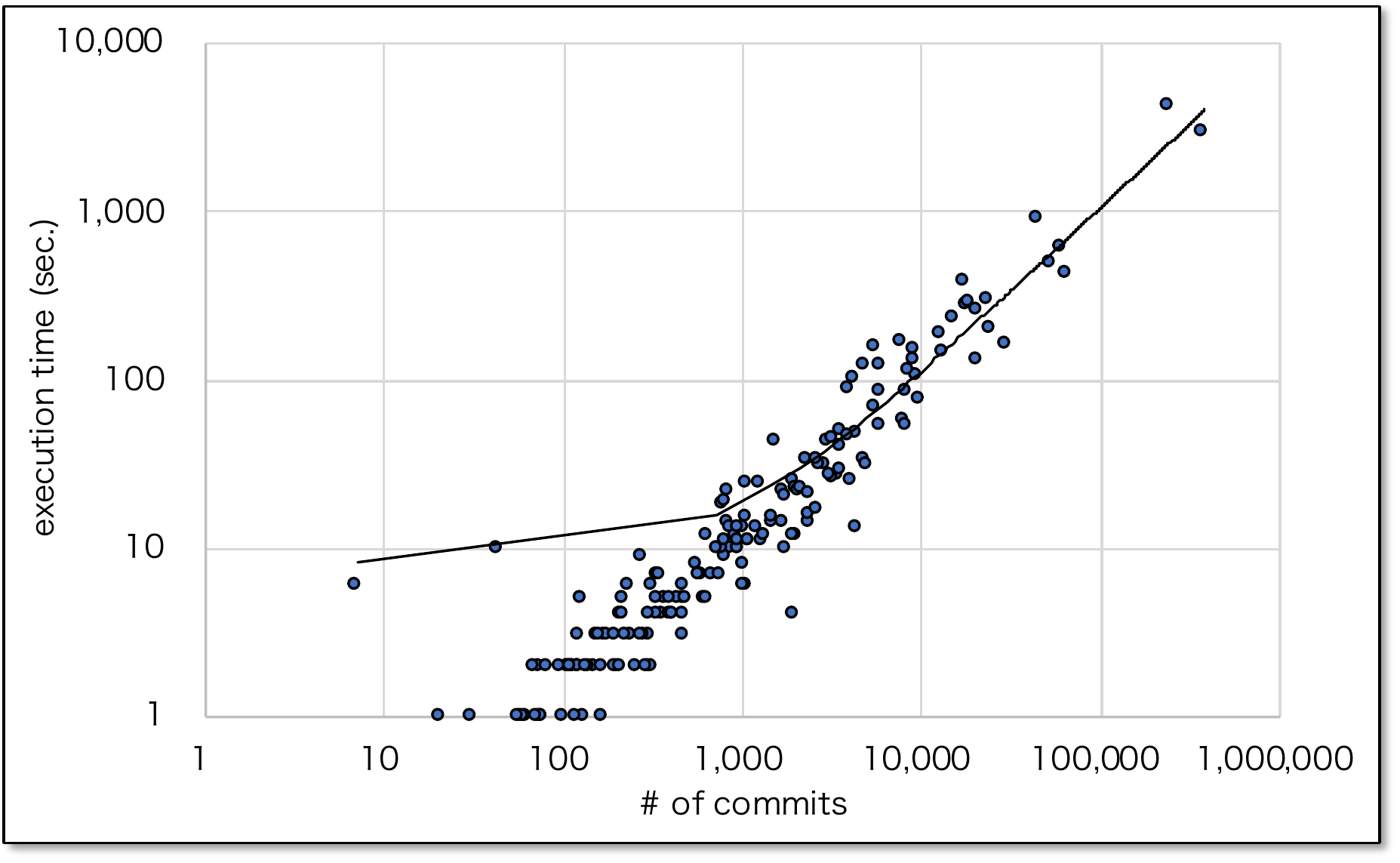}
  \caption{Execution time of \FinerGit.}
  \label{fig:ExecutionTime}
  %\vspace{-1.0em}
\end{figure}

We measured the time that \FinerGit reconstructed the repositories of the target projects on MacBook Pro\footnote{CPU: 2.7GHz quad-core Intel Core i7, memory size: 16 GBytes}.
Figure~\ref{fig:ExecutionTime} shows the measurement results.
This figure shows that \FinerGit is scalable enough for large repositories.
In the longest case, \FinerGit took 4,209 seconds to reconstruct the repository of intellij-community, which includes more than 240K commits.
Of course, this execution time \R{\ID{\#2.8}can be shorter} if a higher specification computer is used\footnote{We also measured execution time with our workstation whose CPU is 3.6GHz octet-core Intel Core i9 and memory size is 32 GBytes. The execution time was approximately 22\% of MacBook Pro's one.}.

Figure~\ref{fig:ExecutionTime} includes the regression line for all the data.
The regression line shows that \FinerGit takes around 100 seconds to process each 10K commits for large repositories.

\section{\ID{\#1.2}Comparisons with Other Techniques}
\label{sec:comparison}

We also compared \FinerGit with two other techniques, \AURA and RefactoringMiner~(\RMiner).
The first comparison target is \AURA, which is a technique that takes two versions of Java source code and generates mappings of methods between them~\cite{wu2010icse}.
\AURA performs call dependency and text similarity analyses to generate mappings.
The second comparison target is \RMiner, which is a technique that detects refactorings from commit history~\cite{tsantalis2018icse}.
\RMiner's refactoring detection is based on an AST-based statement matching algorithm.
\RMiner defines different rules for different refactoring patterns.
\RMiner checks if matching results of two ASTs before and after changes in a given commit follow any of the rules.

We conducted this comparison on the development history of JHotDraw between releases 5.2 and 5.3.
This development history is one of the evaluation targets in \textsf{AURA}'s literature~\cite{wu2010icse}.
Releases 5.2 and 5.3 include 1,519 and 1,981 methods, respectively.
There are 19 commits between releases 5.2 and 5.3.

\subsection{AURA}
\label{sec:comparison:aura}

\begin{table}[b]
  \centering
  \caption{Refactorings detected by \RMiner}
  \label{tbl:rminerresults}
  \begin{tabular}{lr} \hline
    Refactoring pattern & \# of detected instances \\\hline
    \Ref{Change Parameter Type}	& 56 \\
    \Ref{Change Return Type} & 10 \\    
%    \Ref{Extract Method} & ?? \\
%    \Ref{Extract \& Move Method} & ?? \\
    \Ref{Move Method}	& 3 \\
    \Ref{Rename Method}	& 44 \\
    \Ref{Rename Parameter} & 45 \\\hline
    Total & 158 \\\hline
  \end{tabular}
\end{table}

We made \FinerGit's repository and tracked the 1,981 methods with 20\% threshold with the command shown in Subsection~\ref{sec:evaluation:accurary}.
The tracking results of 185 methods included renaming and the total number of renaming was 241.
Two of the authors independently examined the tracking results to make oracles.
Each author spent several hours on this task.
The two authors make different oracles for 18 out of the 185 methods.
The authors had a discussion on the 18 methods to obtain consensus for them.
After a one-hour discussion, they got consensus oracles for the 18 methods.
Our consensus oracle includes 161 renamings on 124 methods.

Next, we tracked the 1,981 methods with 50\% threshold, which is the best F-measure threshold in the evaluation in Subsection~\ref{sec:evaluation:accurary}.
As a result, we obtained 161 renamings on 124 methods.
By comparing the tracking results of 50\% threshold with the consensus oracle, 
We calculated two kinds of precision and recall: one was calculated based on renaming instances; the other was calculated based on methods whose tracking results included at least one renaming in the consensus oracle.
\begin{itemize}
  \item From the viewpoint of renaming instances, precision and recall were 91.30\% and 83.52\%, respectively.
  \item From the viewpoint of methods including renames, precision and recall were 86.29\% and 83.59\%, respectively.
\end{itemize}
According to \AURA's literature~\cite{wu2010icse}, \AURA generated mappings for 97 rules\footnote{A rule is a mapping group of multiple methods.} and its precision was 92.38\%. 
By comparing those results, we conclude that \FinerGit generated mappings for more methods with slightly-lower precision.

\AURA utilizes text similarity and call dependency to generate mappings while \FinerGit utilizes only text similarity. 
On the other hand, \AURA takes only two versions of source code to generate mappings while \FinerGit utilizes all commits to track methods.
Those are the reason why the precision values of the two tools were not so different.

\subsection{RefactoringMiner}

We performed \RMiner\footnote{\RMiner is available at \url{https://github.com/tsantalis/RefactoringMiner}. We used the latest version of the tool at 17th November, 2019. The commit ID is \texttt{4bb0e11550b781b61ce1c382a58ea182a2f46944}.} on the commit history of JHotDraw between release 5.2 and 5.3.
\RMiner has a capability of detecting 38 types of refactoring patterns and the following five refactoring patterns correspond to renamings that \FinerGit detects: \Ref{Change Parameter Type}, \Ref{Change Return Type}, \Ref{Move Method}, \Ref{Rename Method}, and \Ref{Rename Parameter}.
\RMiner detected 158 refactoring instances of the five patterns.
The detail numbers of refactorings detected by \RMiner are shown in Table~\ref{tbl:rminerresults}.
We compared the 158 refactorings with the 161 renamings detected by \FinerGit with 50\% threshold.
The number of common instances was 65, which was 41.14\% of \RMiner's refactorings and 40.37\% of \FinerGit's renamings.

\begin{table}[b]
  \centering
  \caption{Precision and Recall of \textsf{RMiner} in literature~\cite{tsantalis2018icse}}
  \label{tbl:accuracy2018icse}
  \begin{tabular}{lcc} \hline
    Refactoring pattern & Precision & Recall \\\hline
%    \Ref{Extract Method} & 98.63\% & 84.72\% \\
%    \Ref{Extract \& Move Method} & 95.92\% & 41.23\% \\
    \Ref{Move Method} & 95.17\% & 76.36\% \\
    \Ref{Rename Method} & 97.78\% & 83.28\% \\\hline
  \end{tabular}
\end{table}

The \FinerGit evaluation in Subsection~\ref{sec:comparison:aura} shows that \FinerGit's tracking accuracy on JHotDraw is high (precision and recall are 91.30\% and 83.52\%, respectively in 50\% threshold).
Table~\ref{tbl:accuracy2018icse} shows precision and recall of \RMiner for each refactoring pattern in literature~\cite{tsantalis2018icse}\footnote{\Ref{Change Parameter Type}, \Ref{Change Return Type}, and \Ref{Rename Parameter} were not investigated in the literature because those refactoring patterns have been recently supported by \RMiner.}.
According to this table, precision and recall of \RMiner are also high.
However, the common instances between \FinerGit and \RMiner do not occupy a large portion of all instances detected by either of the techniques.
We manually investigated renames and refactorings that had been detected only either of the techniques and found that the results faithfully reflected their different inheritances.
There were two major cases of renames that were detected only by \FinerGit.
\begin{itemize}
  \item New parameters were added to methods or return types of methods were changed according to the changes in method's bodies. 
  Those changes were not refactorings but functional enhancements.
  \item Access modifiers (\texttt{public}, \texttt{protected}, and \texttt{private}) were added/removed/changed. 
  Such changes were refactorings; however they were not supported by \RMiner.
\end{itemize}
On the other hand, refactorings that were detected only by \RMiner had changed a large part of method's bodies.
Thus, line similarities of method's bodies between such refactorings become low, which leaded to fail to be detected as a renaming by \FinerGit.

Herein, we compared \FinerGit with \RMiner; however their purposes are different from each other.
The \FinerGit's purpose is tracking Java methods with high accuracy.
No matter what kinds of changes are made, \FinerGit is able to track methods if a line similarity of the method's bodies between a change is higher than a given threshold.
On the other hand, the purpose of \RMiner is detecting refactorings in a commit history.
No matter how unsimilar between method's bodies are between a refactoring, \RMiner is able to detect the refactoring if the refactoring is supported by \RMiner.

\section{Threats to Validity}
\label{sec:validity}

In the experiment, we used 182 Java projects, and we investigated on tracking results on 1,768K methods in total.
Those numbers of projects and methods are large enough so that we expect that the same results are obtained if we conduct another experiment on different Java projects.

To measure precision, recall, and F-measure of method tracking by \FinerGit and \Historage, we manually constructed oracle for 182 methods.
Firstly, two of the authors made oracle for all the 182 methods independently, and then they discussed for which they made different oracle.
This process of making oracle is designed to avoid making mistakes and to reduce subjective view on constructing oracle as much as possible.

One more thing about oracle is that, essentially, oracle should be made independently from tracking results of \FinerGit and \Historage.
However, constructing oracle with a fully-manual work is extraordinarily difficult even for a small number of methods.
Consequently, in the experiment, we firstly obtained high-recall tracking results with an enough low threshold, and then, we checked how many false positives were included in the tracking results.
We consider that this construction process does not ensure 100\%-correct oracle but high enough for comparing different techniques.
In other word, we made oracle of reasonable quality with a realistic time cost.

\R{
\ID{\#2.6}
In the manual investigation, we checked surrounding 15 lines (as shown in Subsection~\ref{sec:evaluation:accurary}) of changes in commits to judge whether method tracking by \FinerGit was correct or not. 
The number 15 came from our experiences with \FinerGit because we had checked tracking results of \FinerGit before conducting the experiment in this paper.
}

In the experiment, we discussed the comparison results by focusing on whether \FinerGit had found more renaming and copying for Java methods than \Historage.
However, we also need to see the fact that there were some cases that short tracking results by \FinerGit were better than long tracking results by \Historage.
Such cases mean that \FinerGit was able to avoid tracking methods incorrectly.
We investigated some of such cases, and then we found that the reason why \Historage found a higher number of renames is due to the existences of coincidentally matched lines as shown in Figure~\ref{fig:Heuristic1}\subref{fig:Heuristic1Historage}.

\section{Related Work}
\label{sec:relatedwork}

The research that is most related to this paper is of course \Historage~\cite{hata2011iwpse}.
\Historage is useful in research on mining software repositories because researchers can obtain Java method histories without implementing code/scripts by themselves.
\Historage has been used in many research before now.
\begin{itemize}
  \item Hata et al.\ researched predicting fault-prone Java methods by using method histories obtained with \Historage~\cite{hata2012icse}.
  Their experimental results showed that the method-level prediction outperformed package-level and file-level predictions from the viewpoint of efforts for finding bugs.
  \item Hata et al.\ also used \Historage to infer restructuring operations on the logical structure of Java source code~\cite{hata2011iwesep}.
  \item Fujiwara et al.\ developed a hosting service of \Historage repositories, Kataribe\footnote{\url{http://sdlab.naist.jp/kataribe/}}~\cite{fujiwara2014msr}.
  Kataribe enables researchers/practitioners to browse method histories on the web, and they can clone \Historage repositories in Kataribe into their local storages if they want to conduct further analyses.
  \item Tantithamthavorn et al.\ investigated the impact of granularity levels (class-level and function-level) on a feature location technique~\cite{tantith2014apres}.
  The results indicated that function-level feature location technique outperforms class-level feature location technique.
  Moreover, function-level feature location technique also required seven times less effort than class-level feature location technique to localize the first relevant source code entity.
  %In that study, they used \Historage to obtain function-level change histories.
  \item Kashiwabara et al.\ proposed a technique to recommend appropriate verbs for a method name of a given method so that developers can use various verbs consistently~\cite{kashiwabara2015ieice}.
  Their technique recommends candidate verbs by using association rules extracted from existing methods.
  They extracted renamed methods from repositories of target projects using \Historage.
  \item Oliveira et al.\ presented an approach to analyze the conceptual cohesion of the source code associated with co-changed clusters of fine-grained entities~\cite{oliveira2015bsse}.
  They obtained change histories of Java methods with \Historage.
  By using the change histories, they identified a set of methods that were frequently changed together.
  \item Yamamori et al.\ proposed to use two types of logical couplings of Java methods for recommending code changes~\cite{yamamori2017compsac}.
  The first type is logical couplings that are extracted from code repositories.
  They used \Historage and Kataribe to obtain logical couplings of Java methods.
  The second type is logical couplings that are extracted from interaction data.
  They used a dataset that had been collected by Mylyn~\cite{kersten2005aosd}.
  Their experimental results showed that there was a significant improvement in the efficiency of the change recommendation process.
  \item Yuzuki et al.\ conducted an empirical study to investigate how often change conflicts happen in large projects and how they are resolved~\cite{yuzuki2015swan}.
  In their empirical study, they used \Historage to conduct method-level analysis.
  As a result, they found that 44\% of conflicts were caused by changing concurrently the same positions of methods, 48\% is by deleting methods, and 8\% is by renaming methods. They also found that 99\% of the conflicts were resolved by adopting one method directly.
  \item Suzuki et al.\ investigated relationships between method names and their implementation features~\cite{suzuki2017bcd}.
  They showed that focusing on the gap between method names and their implementation features is useful to predict fault-prone methods.
  They used \Historage to collect change histories of Java methods in the investigation.
\end{itemize}
All the above research can be conducted with \FinerGit instead of \Historage.
Moreover, the experimental results may change if \FinerGit is used because there is a significant difference in the tracking results between \FinerGit and \Historage.

We are not the first research group that has used single-token-per-line format for \Git repositories.
\R{\ID{\#2.8}To the best of our knowledge, the study by German et al.\ was the first attempt to follow this approach~\cite{german2019underreview}.}
They proposed to rearrange source files with single-token-per-line for enabling fine-grained \texttt{git-blame}.
By using their technique, we can see the person who changed last for each token of the source code.
They showed that blame-by-token reports the correct commit that adds a given source code token between 94.5\% and 99.2\% of the times, while the traditional approach of blame-by-line reports the correct commit that adds a given token between 74.8\% and 90.9\%.
German developed a system \cregit\footnote{\url{https://github.com/cregit/}} based on their proposed technique.
\cregit has being used in Linux development community\footnote{\url{https://cregit.linuxsources.org/}}.
\cregit does not extract Java methods as files, which is a difference between \cregit and \FinerGit.

\R{\ID{\#1.1}
Heuristic-1, which is described in Subsection~\ref{sec:heuristics}, is refining symbols in source code.
On the one hand, symbol refinements are often performed in the process of code clone detection techniques.
In the context of clone detection, some symbols are replaced with special ones prior to the matching process.
For example, in \Tool{CCFinder}~\cite{kamiya2002tse} and \Tool{NICAD}~\cite{roy2008icpc}, which are representative code clone detection techniques, all variables and literals are replaced with a specific wildcard symbol.
The purpose of replacements is to detect syntactically-similar code as code clones as much as possible.
Such replacements can realize that the matching process ignores differences in variables or literals.
On the other hand, in the context of \FinerGit, we do not want to ignore differences in variables or literals.
If we ignore such differences, the similarity between non-related methods can rise accidentally, which leads \FinerGit to make wrong method tracking. 
The purpose of our Heuristic-1 is to calculate lower similarity values between non-related methods.
}

There are many research studies of program element matching other than \Historage~\cite{godfrey2005tse}.
Lozano et al.\ and Saha et al.\ implemented method tracking techniques since they need to track method-level clones in their experiments~\cite{lozano2008icsm,saha2011icsm}.
Their method-level tracking techniques are line-based comparisons and their comparisons compute numerical similarity values by comparing lines as texts.
Thus, in the case that only a small part of a line is changed, the similarity between a before-change line and its after-change line should be high while a simple line-based comparison like \texttt{diff} regards that a before-change line is completely different from its after-change line.
However, their comparisons are still line-based ones, which include some flaws compared to token-based ones.
\begin{itemize}
  \item In the cases that the first token of the line is moved to the previous line or the last token of the line is moved to the next line (e.g., left bracket (``\{'') is moved to the next line due to format change), their line-based techniques regard that multiple lines have been changed while our technique regards that no lines have been changed.
  \item The same changes have different impacts on lines of different length. For example, variable \texttt{abc} is changed to \texttt{def} in a 10-character line, the similarity becomes 7/10 while the same change occur in a 40-character line, the similarity becomes 37/40.
\end{itemize}

Godfrey and Zou detected merging and splitting source code entities such as files and functions.
They extended origin analysis~\cite{tu2002iwpc} to track source code entities.
They utilize various information for entities such as entity names, caller/callee relationship, and code metrics values.
Wu et al.\ proposed a technique to identify change rules for one-replaced-by-many and many-replaced-by-one methods~\cite{wu2010icse}.
Their approach is a hybrid one, which means that it uses two kinds of data: caller/callee relationship and text similarity.
Kim et al.\ proposed a technique to track functions even if their names get changed~\cite{kim2005wcre}.
Their technique computes function similarities between given two methods.
They introduced eight similarity factors such as complexity metrics and clone existences to determine if a function is renamed from another function.
Dig et al.\ proposed a technique to detect refactorings performed during component evolution~\cite{dig2006ecoop}.
Their technique can track methods even if refactorings change their names.
Their detection algorithm uses a combination of a fast syntactic analysis to detect refactoring candidates and a more expensive semantic analysis to refine the results.
There are many other approaches for identifying refactorings, and many of them support refactorings that changes method names/signatures such as \Ref{Rename Method} and \Ref{Parameterize Method} pattern~\cite{kim2010fse,milea2014fse,prete2010icsm,silva2017msr,tsantalis2018icse,weissgerber2006ase,xing2005ase}.
The advantage of the proposed technique against the above approach should be the ease to use because it utilizes \Git mechanisms to track methods.
A researcher/practitioner who wants method evolution data does not have to learn how to use new tools.

\section{Conclusion}
\label{sec:conclusion}

In this paper, we firstly discuss \Historage, which is proposed in literature~\cite{hata2011iwesep}.
\Historage is a tool that converts a \Git repository to a finer-grained one.
In the finer-grained repository, each Java method exists as a single file.
Thus, we can track Java method with \Git commands such as \texttt{git-log}.
However, tracking small methods with \Git mechanisms does not work well because small methods do not have good chemistry with the \Git rename detection function.
Thus, we proposed a new technique that puts only a single token of Java methods per line.
In other words, in our technique, each line includes only a single token.
We also derived two heuristics to reduce incorrect tracking.

We implemented a software tool based on the proposed technique.
We applied our tool and \Historage to 182 repositories of Java OSS projects to compare the two tools.
The 182 repositories include 1,768K methods in total, which are the targets our comparisons.
We found that \FinerGit scored 84.52\% as maximum F-measure while \Historage scored 70.23\%.
We also confirmed that the proposed technique worked well for methods of any size in spite that our research motivation was to realize better tracking for small methods.
Furthermore, we showed that our tool took only short time to construct finer-grained repositories even for large repositories.

In the future, we are going to replicate some experiments of existing research with \FinerGit to check whether the better tracking of our tool changes experimental results or not.

\section*{Acknowledgements}

This work was supported by JSPS KAKENHI Grant Number JP17H01725 and JP18K11238.

\bibliographystyle{elsarticle-num}
\bibliography{references}

\end{document}